\documentclass[12pt]{iopart}

 \usepackage{transparent}
\usepackage{floatrow}
\usepackage{cite}
\usepackage[colorlinks=true,breaklinks,
]{hyperref} 
\usepackage[hyphenbreaks]{breakurl} 
\usepackage{xcolor}
\definecolor{internal}{rgb}{0.6,0,0.2} 
\definecolor{citationC}{rgb}{0,0,0.6} 
\definecolor{c3}{rgb}{0.3,0,0.9} 
\hypersetup{
    linkcolor= {internal}, 
    citecolor={citationC}, 
    urlcolor={c3} 
}
\usepackage{color}
\usepackage{url}
\usepackage{booktabs}
\usepackage{float}
\usepackage{amssymb}
\usepackage{amsthm}
\usepackage{graphicx}
\usepackage{subfig}
\usepackage[space]{grffile}
\usepackage{tensor}
\usepackage{amsfonts}
\usepackage{bm}
\usepackage[utf8]{inputenc}
\usepackage{slashed}
\usepackage{hyperref}
\usepackage[toc,page]{appendix}	
\usepackage{lipsum} 
\usepackage{comment}
\usepackage[percent]{overpic}
\usepackage{tikz}
\usetikzlibrary{arrows.meta}
\usetikzlibrary{bending}
\usetikzlibrary{decorations.text}

\newcommand{\ext}[1]{\mathbf{d}\ensuremath{\mathbf{#1}}}
\providecommand{\abs}[1]{\vert#1\vert}

\def\beq{\begin{eqnarray}}
\def\eeq{\end{eqnarray}}

\definecolor{axisColor}{rgb}{0, 0, 0.5}

\begin{document}

\title[]{Collins in Wonderland}

\author{Ben David Normann$^1$, Sigbj\o rn Hervik$^1$}


\address{$^1$Faculty of Science and Technology, University of Stavanger, 4036, Stavanger, Norway\\
}
\ead{bdnormann@gmail.com, sigbjorn.hervik@uis.no}
\vspace{10pt}
\begin{indented}
\item[]February 2019
\end{indented}

\begin{abstract}
What is the asymptotic future of a scalar-field model if the assumption of isotropy is relaxed in generic, homogeneous space-times with general relativity? This paper is a continuation of our previous work on Bianchi cosmologies with a $p$-form field (where $p\,\in\,\{1,3\}$)---or equivalently: an inhomogeneous, mass-less scalar gauge field with a homogeneous gradient. In this work we investigate such matter sector in General Relativity, and restrict to space-times of the particular Bianchi types VI$_0$ and VI$_{\tilde{h}}$, where $\tilde{h}=h<0\,\cap\,\neq\,-1/9\,\cup\,-1$. We show that the previously found fabric of exact solutions named Wonderland are future attractors in $\mathcal{B}$(VI$_0$) and $\mathcal{B}$(VI$_{\tilde{h}}$), extending the Collins perfect-fluid equilibrium set to include a $p$-form (with $p\,\in\,\{1,3\}$). We also write down the line-element corresponding to Wonderland in VI$_{\tilde{h}}$ and give explicit expressions for the underling gauge-potential $\phi(t,\mathbf{x})$ corresponding to this solution. Simulation of a path approaching Wonderland in Bianchi type I is also given.
\end{abstract}
%
\vspace{2pc}

\section{Introduction}
\label{Intro}
The work contained in this paper may be viewed as a (phenomenological) study of isotropy-breaking dark matter models in anisotropic backgrounds, as continued from~\cite{normann18,normann20a} and also~\cite{thorsrud20}.

\textit{The concordance model} of cosmology has proved very successful in accounting for cosmological observations. This model, often referred to as the $\Lambda$CDM-model due to its two main energy constituents, \textit{the cosmological constant} ($\Lambda$) and \textit{cold dark matter} (CDM), is an exceedingly simple model built on a maximally symmetric spatial background geometry described locally by the \textit{Friedmann-Lemaître-Robertson-Walker} (FLRW) metric.

The model relies on a wealth of (recent) observational testimony which together puts very tight constraints on model parameters. For instance, one finds according to the Planck Collaboration that $\Omega_{K}=0.0007\pm 0.0019$ at $95\%$ confidence level, when combining CMB and BAO measurements\cite{planck18VI}. As a precaution, it is however necessary to keep in mind that the statistical analysis is typically carried out within the framework of FLRW cosmology. The constraints on observables might therefore relax in the full space of homogeneous cosmologies. As a related example, a recent statistical analysis has challenged the flatness paradigm, suggesting instead that the Universe might be spatially closed~\cite{handley19}. According to~\cite{diValentino19}, the pertaining tension between the observed CMB spectra and the predicted lensing amplitude~\cite{planck18VI} results in the CMB spectra favoring a positive spatial curvature at more than the 99$\%$ confidence level.

Observational experiments like COBE~\cite{Smoot92}, WMAP~\cite{komatsu08, komatsu10, hinshaw12} and Planck \cite{planck15XIII,planck13XVI,planck18VII} also reveal that the CMB is observed isotropic to very high degree of accuracy~\cite{planck18VII}, which severely restricts the shear. One must, however, also here be careful to draw the right conclusions. Shear-free cosmologies with anisotropic background geometry are known, and were discussed decades ago~\cite{mimoso93}, and an exact such solution of the Einstein equations with a physical matter model was presented in \cite{carneiro01}, realized by a mass-less scalar field with an isotropy-violating gradient $\nabla_\mu \varphi$. The uniqueness of this solution was more recently established and discussed in~\cite{thorsrud18}, and explicit calculations showing that light propagation in this space-time will produce an isotropic CMB were presented in~\cite{thorsrud20}. It is noteworthy that this unique, shear-free solution is of Bianchi type VI$_{-1}$ (id est; type III). Furthermore, this solution demonstrates that the isotropy of the CMB severely constrains the \textit{shear} but not necessarily the geometry of the model.
 Also, the presence of several large-angle statistical `anomalies'~\cite{copi10,schwarz15,bull15} should leave the option open, that there is an isotropy-breaking field in the Universe yet to be discovered.
 
Although aesthetically satisfying, the observed degree of symmetry described above raises serious questions, like: \textit{How} did the Universe come to take such an isotropic form? Will it sustain an isotropic mode of expansion? These questions are important, as it is indeed a peculiar thing that the Universe has conspired to take such a symmetric form when the theory itself (General Relativity; GR) is much richer, allowing for anisotropic as well as inhomogeneous cosmologies. After all, these questions also underpin the very popular inflation-paradigm, which seeks to explain the observed isotropy of the early Universe by postulating an initial very rapid phase of expansion. It seems inevitable that one must be able to account for the observed degree of isotropy if one intends to understand the developement of the Universe, both into the past and the future.

As a fact of matter: These issues are made all the more relevant considering our severely restricted knowledge of the so-called dark components of the observable Universe.

To explain the observed isotropy, it is necessary to consider the full space of anisotropic, homogeneous cosmologies; The Bianchi models as well as the Kantowski-Sachs model. This should be clear, as the latter requires assumptions that the former does not. Studies of initial anisotropies have been going on for a long time. Consider for instance~\cite{bok:MacCallum79} by MacCallum or the seminal paper \textit{Why is the Universe isotropic?} by Collins and Hawking~\cite{collins73}, both from the seventies. Non-tilted, perfect fluids (cf. \cite{bok:EllisWainwright} and references therein) as well as tilted perfect fluids~\cite{barrow03, hervik04, coley04, coley05, hervik05, hervik07, hervik08, hervik10} and fluids with vorticity~\cite{hervik06a, hervik06b, coley08} have since been considered. Naturally the relation to different inflationary scenarios has been discussed as well (e.g. \cite{barrow06, hervik11}) and the connection to observations has been investigated to some degree, for instance in~\cite{jaffe06}. Finally, we make mention of the e-book by A.~A.~Coley \cite{bok:Coley}, which provides a comprehensible overview of a large range of studies with a variety of matter sectors.

Because geometry and matter are interrelated through the Einstein equations, a study of anisotropic background geometries, must naturally involve an isotropy-breaking matter sector as well. One most natural way to do so is through a $p$-form action. To this end source-free electromagnetism has already been investigated~\cite{leblanc97, yamamoto12,barrow12}. 

It is, from a mathematical point of view, displeasing that the remaining candidates; the $p$-form action with $p\,\in\,\{1,3\}$ have gone largely unnoticed in the cosmological literature until recently when we and collaborators have considered it systematically~\cite{normann18,normann20a,thorsrud18,thorsrud20}~\setcounter{footnote}{0}\footnote{Note in passing also other works, like the short notice \cite{gruzinov04} and the recent works \cite{almeida19,almeida20}.}. As a result, the general equations for a perfect fluid and a homogeneous, sourcefree $j$-form field (where $j=1,3$) in a cosmological context with general relativity were for the first time written down in \cite{normann18}, through the orthonormal-frame approach. It is worthy of notice that the $j$\,-\,form field was only required homogeneous on the \textit{fieldstrength} level. That is to say; the underlying $(j-1)\,$\,-\,gauge field was not required homogeneous. Consequently, one may view the work as a study of an inhomogeneous, mass-less scalar gauge field with a homogeneous gradient. In the forgoing papers, all the Bianchi invariant sets except those of type VI$_0$ and VI$_h$ have been considered~\footnote{The types VIII and XI have only been considered very briefly in \cite{normann18}, since no spatial components of the $p$-form was allowed for the considered cases $p=1$ and $p=3$.} providing a dynamical systems analysis of the cosmological evolution of such universes.

\textbf{The purpose of the present paper} is to study the invariant sets belonging to type VI$_0$ and VI$_{\tilde{h}}$, where $\tilde{h}=h<0\,\cap\,\neq\,-1/9\,\cup\,-1$, investigating the evolution of a $j$-form model in these classes of cosmologies. Furthermore, particular attention is devoted to the analysis of one of the new fabrics of equilibrium sets found across all the Bianchi sets of Solvable type: the so-called \textit{Wonderland} solutions.

The rest of this paper is structured as follows. In the next section we discuss briefly the Bianchi models, and introduce the orthonormal frame, which will be used throughout. In Section~\ref{Sec:SpecMatter} the matter content of the model is studied, followed by Section~\ref{Sec:PhysMot}, where the (physical) motivation for such a study is provided. In Section~\ref{Sec:dynsys} the Bianchi models of type VI$_0$ and VI$_h$ are discussed from a dynamical-systems point of view. A closer look at the two invariant sets belonging to VI$_{\tilde{h}}$\footnote{Here and throughout we define $\tilde{h}=h<0\,\cap\,\neq\,-1\cup-1/9$.} and VI$_0$ follow in Sections~\ref{Sec:B6} and~\ref{Sec:VI0}, respectively. In Sections~\ref{ref:Anis} we provide a closer scrutiny of the various attractors of the sets, whereafter we proceed to have a closer look at the Wonderland solution in~\ref{Sec:scalarfield}. A few simulations are also provided before we finally conclude in Section~\ref{Sec:Concl}.

We set $c=8\pi{\rm G}=1$ throughout.

\newpage
\section{Geometry and frame}
\label{Sec:GometryAndFrame}
In this work we follow the pionering work~\cite{ellis69} in employing the orthonormal-frame formalism, with the time-like axis directed along the time-coordinate $t$. Using numerical indices $\{0,1,2,3\}$ to index the orthonormal frame and the letters $\{t,x,y,z\}$ to index the corresponding coordinate basis we thus have $\mathbf{e}_0=\partial_0=\partial_t$. As further explained in for instance \cite[Chap. 1]{bok:EllisWainwright} the Bianchi types correspond to distinct Lie algebras. A certain Bianchi type may therefore be studied through the structure coefficients $\tensor{\gamma}{^\lambda_{\mu\nu}}$ corresponding to its Lie algebra. Note that Greek indices are taken to run over all four space-time components, whereas Latin indices $\{a,b,c,\cdots\,,m,n\}$\setcounter{footnote}{0}\footnote{We will stick to these letters to avoid confusion: The letters $\{t,x,y,z\}$ \textit{always} refer to particular components of the coordinate basis.} run over spatial components only. For a complete set of basis vectors $\{\mathbf{e}_\mu\}$, these structure coefficients are defined through
\begin{equation}
    [\mathbf{e}_\mu\,,\,\mathbf{e}_\nu]=\tensor{\gamma}{^\lambda_{\mu\nu}}\mathbf{e}_\lambda.
\end{equation}
Defining hypersurfaces by the orbits of the isometry group, we choose the congruence of observers to be hypersurface orthogonal. In particular, we take the four-velocity $\mathbf{u}$ to be aligned with the time-coordinate; $\mathbf{u}=\partial_t$. The motion is now geodesic ($\dot{u}_a=0$)\footnote{Here and throughout $(\dot{\,})$ denotes derivative with respect to time $t$.} and the congruence irrotational ($\omega_{\mu\nu}=0$). Now employing the fact that $\nabla(\mathbf{u}\cdot\mathbf{e}_i)=0$ one finds upon straight forward algebra that the expansion-tensor in an orthonormal frame is given by $\theta_{\mu\nu}=\tensor{\Gamma}{^0_{\mu\nu}}$~\cite[Chap. 15]{bok:GronHervik}. Since we use an orthonormal frame we have the relation $\Omega_{\mu\nu}=-\Omega_{\nu\mu}$ which allows for expressing  $\tensor{\Gamma}{^\alpha_{\mu\nu}}$ in terms of the structure coefficients $\tensor{\gamma}{^\alpha_{\mu\nu}}$. This results in expressions for the mixed structure coefficients $\tensor{\gamma}{^a_{0b}}$ of the orthonormal frame, as given below. The spatial coefficients $\tensor{\gamma}{^c_{ab}}$ are however as usual decomposed according to the so-called Behr decomposition. All in all we have
\begin{eqnarray}
\label{strcoeff}
\tensor{\gamma}{^a_{0b}}=-\tensor{\sigma}{_b^a}-H\tensor{\delta}{_b^a}-\tensor{\varepsilon}{^a_{bm}}\Omega^m,\\
\tensor{\gamma}{^c_{ab}}=\tensor{\varepsilon}{_{abm}}n^{mc}+a_a\tensor{\delta}{_b^c}-a_b\tensor{\delta}{_a^c}\label{behr},
\end{eqnarray}
where $n^{ab}$ is symmetric and trace-free. One also finds that the remaining structure coefficients vanish; $\tensor{\gamma}{^0_{0a}}=\dot{u}_a=0$ and $\tensor{\gamma}{^0_{ab}} =-2\tensor{\varepsilon}{_{ab}^m}\omega_m=0$. In the above, $H$ is the expansion tensor, $\varepsilon_{abc}$ is the totally antisymmetric symbol, $\delta$ is the Kronecker-delta and $\Omega^m$ give the rotations of the frame, defined such that $\Omega^\alpha\,\equiv\,-\frac{1}{2}\varepsilon^{\alpha\beta\gamma\delta}u_\beta\mathbf{e}_\gamma\cdot\mathbf{\dot{e}}_\delta.$
The Jacobi identity must be fulfilled for all members of a Lie algebra. Taking the Jacobi identity for the triple ($\mathbf{e}_a,\mathbf{e}_b,\mathbf{e}_c$) implies that the vector $\mathbf{a}$ lies in the kernel of the matrix $n^{ij}$;
\begin{equation}
n^{ij}a_j=0.
\end{equation}
The Jacobi identity for the triple ($\mathbf{u},\mathbf{e}_a,\mathbf{e}_b$) provides evolution equations for the structure coefficients. In particular, with $\mathbf{u}=\partial_t$, we find
\begin{eqnarray}
&\dot{a}_i=-\frac{1}{3}\theta a_i-\sigma_{ij}a^j+\varepsilon_{ijk}a^j\Omega^k\,,\\
&\dot{n}_{ab}=-\frac{1}{3}\theta n_{ab}+2n^k_{(a}\varepsilon_{b)kl}\Omega^l+2n_{k(a}\sigma_{b)}^k.
\end{eqnarray}
The different invariant sets of the system of evolution and constraint equations obtained through the Jacobi identity give rise to the different Bianchi invariant sets of type I--IX, which we will refer to as $\mathcal{B}$(I), $\mathcal{B}$(II), ... ,$\mathcal{B}$(IX). Without loss of generality~\cite[chapters 1.5~and~1.6]{bok:EllisWainwright}, a choice is made such that $\mathbf{e}_1$ points in the direction of the vector $\mathbf{a}$, leaving the remaining frame vectors $\mathbf{e}_2$ and $\mathbf{e}_3$ defined up to a rotation. We shall adopt the choice 
\begin{equation}
\label{a}
\mathbf{a}=(a,0,0)\quad\quad\quad\quad\textrm{1+1+2 decomposition}.
\end{equation} 
As a consequence, the equations for $\dot{a}_2$ and $\dot{a}_3$ immediately imply
\begin{equation}
\label{gauge23}
\Omega_A=\varepsilon_{AB}\sigma^{1B}\quad\quad\textrm{and}\quad\quad n^{1i}=0.
\end{equation}
Here and throughout capital letters run over $\{2,3\}$ and $\varepsilon_{AB}$ is the totally antisymmetric symbol with $\varepsilon_{23}\,\equiv\,1$. Note that Eq.~\eref{a} carries no information for models of class A, since here $\mathbf{a}=0$. The gauge choice~\eref{gauge23} may still be made, however, in all class A models that admit a $G_2$ subgroup of isometries. By such, it becomes possible to make this choice for all types except VIII and XI, which do not admit a $G_2$ subgroup of isometries.

By the above equation two of the frame rotations are specified. There remains in this way only one rotational gauge freedom: rotation of the frame around the $\mathbf{e}_1$-axis. In this work we work in the $F$-gauge and the $N_-$-gauge. These gauges are further specified in~\ref{App:Gauge}.

\section{Specifying the matter sector}
\label{Sec:SpecMatter}
We take in this paper the matter sourcing
\begin{equation}
\label{matter}
    \rho=\rho_{\rm pf}+\rho_{j{\rm f}}\quad\quad\quad\textrm{where}\quad\quad\quad,
\end{equation}
where $\rho$ means \textit{energy density} and where subscript `pf' refers to the perfect fluid and `$j$f' refers to the $j$-form fluid. The perfect fluid is assumed to be \textit{non-tilted} and \textit{perfect}. Thus it is assumed to fulfill the equation
\begin{equation}
    \label{pf}
    p_{\rm pf}=(\gamma-1)\rho_{\rm pf}\quad\quad\textrm{where}\quad\quad 0\,\leq\,\gamma\,\leq\,2,
\end{equation}
for its pressure $p_{\rm pf}$ and energy density $\rho_{\rm pf}$. The $j$-form originates from the $p$-form action
\begin{equation}
\label{action}
S=-\frac{1}{2}\int{\mathcal{P}}\wedge\star\mathbf{\mathcal{P}}\,,
\end{equation}
where $\mathbf{\mathcal{P}}$ is a $p$-form constructed by the exterior derivative of a ($p-1$)-form~$\mathcal{K}$. The Bianchi identity and the equations of motion may now be given in the language of exterior calculus by the following two equations.
\begin{eqnarray}
\fl\textbf{$p$-form fluid}\quad\cases{
\ext{\mathcal{P}}=0\quad\quad\rightarrow\quad\quad\nabla_{[\alpha_0}\mathcal{P}_{\alpha_1\cdots\alpha_p]}=0\quad\quad &\textrm{Bianchi Id.}\label{dP}\\
\ext{\star\mathcal{P}}=0\quad\quad\rightarrow\quad\quad\nabla_{\alpha_1}\mathcal{P}^{\alpha_1\cdots\alpha_p}=0\quad\quad &\textrm{Eq. of mot.}.}
\end{eqnarray}
The latter equation implies a source-free field, as dictated by the action. One finds, by inspection, that the system of equations \eref{dP} is the same for $p=1$ as for $p=3$. \textbf{We therefore collectively refer to both these mathematical options as the $j$-form,} where $j\,\in\,\{1,3\}$. Take the $1$-form components $\mathcal{J}_{\alpha}$ to be the $j$-form if $j=1$ and its Hodge dual if $j=3$. We now decompose such that
\begin{equation}
\label{J}
\mathcal{J}_\alpha=-w\, u_\alpha+v_\alpha\,,
\end{equation}
where the 4-velocity $u_\alpha$ is time-like ($u_\alpha u^\alpha\,<\,0$), whereas $v_\alpha$ is defined to be orthogonal to $u_\alpha$ and therefore space-like ($v_\alpha v^\alpha\,>\,0$). Then the equations~\eref{dP} give
\begin{eqnarray}
\label{fluid_dstJ}
& \dot{v}_1=-(H+\sigma^{11})v_1-2v_2\sigma^{12}-2v_3\sigma^{13},\\
&v_2=-(H+\sigma^{22})v_2+(\Omega_1-\sigma^{23})v_3,\\
&\dot{v}_3=-(H+\sigma^{33})v_3-(\Omega_1+\sigma^{23})v_2,\\
&v_2 n^{23}+v_3 n^{33}+v_2 a=0,\\
&v_2n^{22}+v_3 n^{23}-v_3 a=0,\\
&\partial_0w=-3Hw-2v_1 a,
\end{eqnarray}
when using~\eref{a} and~\eref{gauge23} and also invoking the fact that homogeneity requires the vanishing of all spatial derivatives. It is hence clear from the two first equations that if it is possible to choose $v_2=v_3=0$ initially, then these components will remain zero throughout. In the dynamical systems here considered, such a choice is possible, as discussed in~\cite{normann20a}.

Finally, note that all the matter components $i$ are assumed to be non-interacting. Id est, the equation
\begin{equation}
    \label{noninteracting}
    \nabla_\mu \left(T^{\mu\nu}\right)_i=0
\end{equation}
is fulfilled for each energy-momentum tensor $T_i$.
\section{Physical motivation for studying the $j$-form field:}
\label{Sec:PhysMot}
One may wonder, perhaps, how physical the phenomenological study undertaken in this line of work is. Our main argument for why it should be of potential physical interest, is its equivalence to scalar gauge-field models. Take first $\mathcal{P}$ to be a 1-form constructed from an underlying gauge-potential $\phi(\mathbf{x},t)$, such that
\begin{equation}
\label{relapfix}
\ext{\phi}=\mathcal{P}\quad\rightarrow\quad\mathcal{P}_\mu=\partial_\mu\phi.
\end{equation}
Equations ~\eref{dP} now yield
\begin{eqnarray}
\label{dphi}
\ext{\ext{\phi}}&=0\quad\rightarrow\quad\nabla_{[\mu}\nabla_{\nu]}\phi=0\quad\rightarrow\quad\partial_{[\mu}\partial_{\nu]}\phi=\tensor{\gamma}{^\lambda_{\mu\nu}}\partial_\lambda\phi\\
\label{dStphi}\ext{\star\ext{\phi}}&=0\quad\rightarrow\quad\nabla_{\mu}\nabla^{\mu}\phi=0\quad\rightarrow\quad\partial_{\mu}\partial^{\mu}\phi=\tensor{\gamma}{^\mu_{\mu\nu}}\partial^\nu\phi.
\end{eqnarray}
These are the equations for a mass-less scalar gauge field. In a Lagrangian formulation, then, our study is the study of the matter lagrangian
\begin{equation}
\label{1formb}
\mathcal{L}=-\frac{1}{2p!}\partial_\mu\phi\partial^\mu\phi.
\end{equation}
This one can also see directly from the action \eref{action} (with $p=1$) through the definition $S=\int\sqrt{g}\mathcal{L}_\phi{\rm d}^4x$. 
Scalar-field models of this kind have received overwhelming attention. For a review, refer for instance to the recent review~\cite{lopez19}. As such, the current line of work might be seen as an effective means to casting such theories in anisotropic backgrounds, to analyse the effects of such matter sourcing in the full theory of homogeneous cosmologies with GR. What are the effects on such cosmologies when the requirement of isotropy is relaxed?
Hence, \textbf{our study is equivalent to the study of a mass-less, inhomogeneous scalar gauge-field with a homogeneous gradient} in anisotropic background geometry.

Morover, it is important to note that the $p$-form action obeys the weak energy condition. From a Hamiltonian point of view, it is bounded from below, as shown explicitly in \cite[Sec. 2.3]{thorsrud18}. The $j$-form field cannot sustain an accelerated state of expansion, but even so it may still play an important role in the early universe. Isotropy-breaking fields of all sorts have previously been investigated in this context \cite{hervik11,ford89,ackerman07,ackerman09,golovnev08,watanabe09,maleknejad13,ito15, almeida19,cicciarella19}, motivated as they often are by addressing the $\Lambda$CDM anomalies\cite{bennett11}. The $p$-form field is a simple way to incorporate an isotropy-breaking field in a general manner.

\subsection{Diverse further physical motivation of the more speculative kind}
Interpreting the $j$-form as a $3$-form, one may note that the strong-CP problem is precisely due to a mass-less $3$-form field, which, from observational bounds, must be truly minuscle\setcounter{footnote}{0}\footnote{According to Prof. G.~Dvali $F\,<<\,10^{-10}$ (QCD units).}. This suggests that cosmological scenarios including such fields alongside a mechanism for driving them to zero, should be of potential physical interest. For a discussion, consider for instance \cite{dvali05}.

Moreover, it is noteworthy that the effective axion fieldstrength may be described by a 1-form. The axions are hypothetical elementary particles that could resolve the strong CP-problem of Quantum Chromo Dynamics (QCD) through the so-called Peccei--Quinn mechanism. Observationally viable axions are very light, but not mass-less, as discussed already long time ago \cite{preskill83,abbott83,dine83}. Since it is light, but not mass-less, and because it resolves the strong CP-problem, the axion seems to be a good candidate for CDM, as for instance noted in the extensive particle physics review~\cite[Sec. 111.6]{review18}. In 2016 the mass of the axion was found to be $0.05\,{\rm meV}\,<\,m_{\rm a}\,<\,1.5\,{\rm meV}$ based on simulations of the early, post-inflationary Universe \cite{borsanyi16}. Judging from the number of citations, the paper obviously draws attention. Recently, another paper was published \cite{braaten19}, investigating how axion-stars should form as a Bose--Einstein condensate, if indeed axions are a major constituent of~CDM. According to \cite{braaten19} the Lagrangian for the axion becomes
\begin{equation}
    \label{axionL}
    \mathcal{L}=\frac{1}{2}\partial_\mu\phi\partial^\mu\phi+V(\phi),
\end{equation}
at moment scales below the confinement-scales of QCD (1~GeV). Here $\phi$ is a real scalar field, and the potential $V(\phi)$ is the self-interaction of the axions from their coupling to gluons. The mass $m_{\rm a}$ of the axion is now given by $V''(\phi)=m_{\rm a}$. In this work we have studied the mass-less case by choosing $V(\phi)=0$. To further explore the cosmological implications of the axion as CDM it would therefore be interesting in the future to extend the scope to models where $m_{\rm a}\,>\,0$.

Next, we could turn to string-theory which provide a plethora of options for $p$-forms and one could expect such to appear in effective low energy actions. A $3$-form fieldstrength was investigated by Barrow and Dabrowski in \cite{barrow97b} in a Kantowski--Sachs background geometry. Barrow et al. also determined the number of independent components possible in a spatially homogeneous cosmological model in dimension 1+3 in the context of string cosmology, with a $3$-form fieldstrength built from a purely time-dependent $2$-form potential \cite{barrow97a}. Also, the study was undertaken in an orthonormal frame. In our case, however, we allow the underlying potential to vary both with space and time.

The  seminal paper by Kalb and Ramond on cosmic strings \cite{kalb74} and action at a distance must also be mentioned. Note that equation~(3.22) therein is precisely the action we start from.

As a last cosmological application---one whose relevance became particularly clear in the light of LHC seeing no new physics---we mention that a 3-form field is also a candidate for solving the Hierarchy Problem by the cosmological relaxation of the Higgs mass via a 3-form field. This scenario was introduced by Dvali and Vilenkin in~\cite{dvali04, dvali06} and more recently  discussed in~\cite{dvali19}.

\vspace{5em}
\section{The dynamical systems of Bianchi type VI}
\label{Sec:dynsys}
In this section we will study two of the invariant subsets with Lie algebras of Bianchi type VI. Dynamical-systems theory is used, and the past- and future attractors found. The reason for studying the system of equations in this way, is that the equilibrium sets, which typically correspond to assymptotic states of the variables, correspond to self-similar, exact solutions of GR (alongside the matter equations). 


Since we are interested in self-similar cosmological models, it is convenient to work in expansion-normalized variables. By such, the time-dependence resulting from the self-similar expansion is `factored out' of the system. It is instructive at this point to have a look at the expansion-normalized variables used. These are given in the following listing, and we refer the reader to \ref{App:CompVar} for relations to the non-normalized variables.
\begin{itemize}
\item \textbf{Matter: } The $j$-form~\eref{J} is decomposed such that $\mathcal{J}/(\sqrt{6}H)=(\Theta,V_1, V_2,V_3)$. Also recall that $\mathbf{V}_c=V_2+iV_3$\setcounter{footnote}{0}\footnote{Refer to \ref{App:CompVar}---or even better; to \cite{normann18}---for details and further definitions.} Furthermore, $\Omega_{\rm pf}$  is the energy density of the perfect fluid.
\item \textbf{Observers: }$\mathbf{\Sigma}_1$,$\mathbf{\Sigma}_\Delta$,$\Sigma_+$ represent the shear of the congruence of observers, here chosen to be co-moving with the perfect fluid.
\item \textbf{Geometry: }$\mathbf{N}_\Delta$, $N_+$ and $A$ describe the curvature of the spatial 3-surfaces.
\item \textbf{Frame: }The quantity $R_1$ (alongside an initial angle $\phi_1$) represents the gauge freedom left in choosing the rotation (and initial orientation) around the $\mathbf{e}_1$ -axis of the orthonormal frame.
\end{itemize}
In the following sections we will study the euquilibrium sets in more detail. Some of the variables will show up as parameters of the various equilibrium sets. When so, we use the following notation.
\begin{eqnarray}
    N_+=\nu_1\quad,\quad N_-=\nu_2\quad,\quad N_\times=\nu_3\quad\textrm{and}\quad\nu^2=\nu_2^2+\nu_3^2,\\
    \Sigma_+=\beta_1\quad,\quad\Sigma_-=\beta_2\quad,\quad\Sigma_\times=\beta_3\quad\textrm{and}\quad \beta^2=\beta_1^2+\beta_2^2+\beta_3^2.
\end{eqnarray}
Note that it is not the derivatives of the variables themselves that vanish on the equilibrium sets. Rather, the derivatives of \textit{scalars}---the gauge independent quantities---must vanish~\cite{coley05}. This means that those quantities that are the same no matter how the orthonormal frame rotate around the $\mathbf{e}_1$-axis, must remain constant. In particular
 $\nu^2\,=\,\bm{N}_\Delta\bm{N}_\Delta^*$ and $\sigma^2\,\equiv\,\bm{\Sigma}_\Delta\bm{\Sigma}_\Delta^*=\beta^2-\beta_1^2$\setcounter{footnote}{0}\footnote{Our notation is somewhat cumbersome at this point: $\beta^2$ is not gauge independent, but $\nu^2$ is. We keep it this way, however, to connect with the notation used in previous works, and to keep the notation in the tables at a minimum.} are such scalars. The same is true for the complex scalar ${\delta^2\,\equiv\,\bm{N}_\Delta\bm{\Sigma}_\Delta^*=\nu_2\beta_2+\nu_3\beta_3+i(\nu_3\beta_2-\nu_2\beta_3)}$. We may thus find that the equilibrium sets possess evolving $\Sigma_-,\Sigma_\times,N_-, N_\times$, as long as $\sigma,\nu,\delta$ remain constant on the motion.
 \vspace{5em}
\subsection{The different invariant sets}
The dynamical system considered in this work is explicitely given in~\ref{App:dynsys}. The overall group constraint that holds for all these models is 
\begin{equation}
\label{grConstr}
\fl \mathcal{B}({\rm VI}):\quad\,\abs{\mathbf{N}_\Delta}^2-N_+^2/3\,>\,0.
\end{equation}
Also, each value of the group parameter $h$ (defined in \eref{h}) corresponds to an invariant set. In this paper we study {$\tilde{h}=\{h<0\,\cap\neq-1\cup-1/9\}$} collectively. We exclude however the special case $h=-1$ (which we have already studied, cf. ~\cite{thorsrud20}) and the exceptional case $h=-1/9$. In the following we provide a complete list of the various invariant subsets with corresponding Lie algebra of Bianchi type VI.
\begin{itemize}
\item $\mathcal{B}$(VI$_{\tilde{h}}$): Here $h\,<\,0\,\cap\,\neq\,-1\,\cup\,-1/9$: In this case, $A\,\neq\,0$.
\item $\mathcal{B}$(VI$_0$): Since $h=0$, eq. \eref{h} shows that $A=0$. The dynamical system for this case may therefore be found by enforcing $A=0$ in $\mathcal{B}$(VI$_h$).
\item $\mathcal{B}$(VI$_{-1}$), also called $\mathcal{B}$(III): Here $A\,\neq\,0$. This is the special case of $\mathcal{B}$(VI$_h$) which needs a different treatment because it allows for an extra degree of freedom of the $j$-form field. The dynamical system for this case may therefore \textit{not} be obtained by enforcing $h=-1$ in $\mathcal{B}$(VI$_{\tilde{h}}$). This extra degree of freedom provides interesting features, and a recent study shows that this is the only Bianchi type in which a shear-free solution with a lower-bounded Hamiltonian exists for a matter sector constructed from $p$-form gauge fields \cite{thorsrud18}.
\item $\mathcal{B}$(VI$_{-1/9}$): Here $A\,\neq\,0$. This special case allows for an extra shear degree of freedom, and the dynamical system for this case may \textit{not} be obtained by enforcing $h=-1/9$ in $\mathcal{B}$(VI$_{\tilde{h}}$). It is called the exceptional case, denoted $\mathcal{B}$(VI$^*_{-1/9}$), whenever this extra degree of freedom is included. Chapter 8 in \cite{bok:EllisWainwright} and also \cite{hervik08} deal with the non-tilted and tilted cases, respectievely.
\end{itemize}
In the following we will provide an analysis of the invariant sets $\mathcal{B}$(VI$_{\tilde{h}}$) and $\mathcal{B}$(VI$_0$) described above. The set $\mathcal{B}$(VI$_{-1}$) is as mentioned already (partially) studied, and $\mathcal{B}$(VI$_{-1/9}$) is left out because of its complexity. The actual dynamical system that we will study was derived in \cite{normann20a} and given in Sec. 4.1 therein. As already mentioned, it is also given in this paper's \ref{App:dynsys}\setcounter{footnote}{0}\footnote{The cases $h=-1/9$and $h=-1$ may not be studied from this dynamical system, as discussed more properly in~\cite{normann20a}.}. Referring to the system of equations found therein, one may observe that in the two sets we here intend to study, one may always, without loss of generality, choose $\mathbf{\Sigma}_\Delta=\mathbf{V}_{\rm c}=0,$
as further discussed in the mentioned source.


\subsection{General behaviour of universes belonging to $\mathcal{B}($VI$_h$) and $\mathcal{B}$(VI$_0$) } 
In \cite[Thrm. 6.2]{normann18} we proved that Bianchi invariant sets with the matter content of Eq.~\eref{matter} where the perfect fluid is restricted to $0<\gamma<2/3$ becomes quasi-de sitter  with $q=\frac{3}{2}-1$. The interesting $\gamma$-range to study in the presence of a $j$-form fluid is therefore $2/3\,\leq\,\gamma\,<\,2$. Since the Lie algebras of type VI$_h$ is of Solvable type, the results just described apply.

We have previously also obtained results in the absence of the perfect fluid. More specifically; we have shown~\cite[Prop 1]{normann20a} that all universes with $A^2>0$ belonging to the invariant sets $\mathcal{B}$(IV), $\mathcal{B}$(V), $\mathcal{B}$(VI$_h$)\footnote{We must here take the opportunity to report a misprint in Proposition 1 in our previous paper~\cite{normann20a}. The set $\mathcal{B}$(VI$_h$) is not mentioned there, but should be in the list.} and $\mathcal{B}$(VII$_h$) and with $\Omega_{\rm pf}=0$ will be past asymptotic to Jacob's Extended Disk (JED) into the past and future asymptotic to Plane Waves (PW). That is to say: It will start off from a vacuum state where only the time-like component of the $j$-form is present and then evolve into a future where the $j$-form is present with a vector mode as well. Refer to Tables~\ref{tab:VIhNm} and~\ref{tab:VIhGg} for specifications of the above mentioned equilibrium sets.

Also for the particular point $\gamma=2/3$ have results already been obtained. Actually, Proposition 2 in ~\cite{normann20a} establishes that all universes belonging to $\mathcal{B}$(IV),  $\mathcal{B}$(VI$_0$), $\mathcal{B}$(VI$_h$), $\mathcal{B}$(VII$_0$) and  $\mathcal{B}$(VII$_h$)  with a perfect fluid and a $j$-form fluid are asymptotically shear-free with $1=\Omega_{\rm pf}+A^2$ and $\Omega_{\rm pf}'=A'=N_+'=0$.

\subsection{Existence of a stable pesent-day anisotropic dark-fluid Universe}
In the previous paragraph we gave results in the case where $\gamma\,\leq\,2/3$, or on the whole $\gamma$-range, but in the absence of the $j$-form fluid. Now; the physically most interesting case as far as present-day cosmologies filled with dark matter is concerned, is naturally $\gamma=1,$ \textit{with} the $j$-form fluid. For $\gamma\,>\,2/3$, we have previously shown that there exists a family of solutions called Wonderland, which has proven to be stable on most of its existence\setcounter{footnote}{0}\footnote{It is stable only for $\gamma\,\in\,(6/5,4/3)$ in type I, and not stable in $\mathcal{B}$(VI$_{-1}$), where another anisotropic attractor is found, cf. ~\cite{thorsrud20}.}, as previously investigated. The stability of $j$-form models in Bianchi type VI$_h$was however not studied in these previous works, and we intend in this paper to show that Wonderland is an attractor solution also in the sets $\mathcal{B}$(VI$_0$) and $\mathcal{B}$(VI$_{\tilde{h}}$). The stability of the different subsets of Wonderland belonging to the various invariant Bianchi types is summarized next.

\subsubsection{The Wonderland fabric, $W(\kappa,\nu_1,\nu^2)$}
This fabric stretches over many Bianchi invariant sets, as discusssed in~\cite{normann20a}, and the overaching specifications of this set is there shown to be
\begin{eqnarray}\fl
\label{Wspec1}
\beta_1=\frac{1}{4}(2-3\gamma)\quad,\quad \beta_2=-\kappa\nu_3\quad,\quad\beta_3=\kappa\nu_2\quad,\quad\nu_1\nu^2=0\\
\fl\label{Wspec2} A=-\kappa(1+\beta_1)\quad,\quad V_1^2=-\beta_1(1+\beta_1)-\nu^2\quad,\quad\Theta=\kappa V_1.
\end{eqnarray}
The family $W(\kappa,\nu_1,\nu^2)$ may be divided into the following subsets.
\begin{itemize}
\item $\mathcal{S}^+(I)\,\supset\,\mathcal{P}_{W}\equiv \lim_{\kappa,\nu_1,\nu\rightarrow 0}W(\kappa,\nu_1,\nu^2)$.
\item $\mathcal{S}^+(V)\,\supset\,\mathcal{P}_{W(\kappa)}\equiv \lim_{\nu_1,\nu\rightarrow 0}W(\kappa,\nu_1,\nu^2)$.
\item $\mathcal{S}^+(\textrm{VII}_h)\,\supset\,\mathcal{P}_{W(\kappa,\nu_1)}\equiv \lim_{\nu\rightarrow 0}W(\kappa,\nu_1,\nu^2)$.
\item  $\mathcal{S}^+(\textrm{VII}_0)\,\supset\,\mathcal{P}_{W(\nu_1)}\equiv \lim_{\kappa,\nu\rightarrow 0}W(\kappa,\nu_1,\nu^2)$.
\item $\mathcal{C}^+(\textrm{VI}_{\tilde{h}})\,\supset\,\mathcal{P}_{W(\kappa,\nu^2)}\equiv \lim_{\nu_1\rightarrow 0}W(\kappa,\nu_1,\nu^2)$.
\item $\mathcal{S}^+(\textrm{VI}_0)\,\supset\,\mathcal{P}_{W(\nu^2)}\equiv \lim_{\kappa,\nu_1\rightarrow 0}W(\kappa,\nu_1,\nu^2)$.
\end{itemize}


\section{The invariant set $\mathcal{B}$(VI$_h$)}
\label{Sec:B6}
The closure of $\mathcal{B}$(VI$_{\tilde{h}}$)is
\begin{equation}
\overline{\mathcal{B}(\textrm{VI}_{\tilde{h}})}=\mathcal{B}(\textrm{VI}_{\tilde{h}})\cup\mathcal{B}(\textrm{VI}_0)\cup\mathcal{B}(\textrm{V})\cup\mathcal{B}(\textrm{IV})\cup\mathcal{C}(\textrm{II})\cup\mathcal{C}(\textrm{I}).
\end{equation}
Hence, we must expect to find equilibrium sets from many other Bianchi invariant sets. Tables \ref{tab:VIhNm} and \ref{tab:VIhGg} provide an overview of the equilibrium sets analysed in $N_-$\,-\,gauge and in $F$\,-\,gauge, respectively.

\begin{table}[H]
	\centering
	\resizebox{\textwidth}{!}{\begin{tabular}{llccccccccccc}
			\toprule
			\multicolumn{13}{c}{\textbf{Equilibrium sets in $\overline{\mathcal{B}(\textrm{VI}_{\tilde{h}}})$ analysed in $N_-$\,-\, gauge.}} \\
			\hline
Bianchi t.&$\mathcal{P}$ & $q$ & $\gamma $&$\alpha^2$&$A$&$\Omega_{\rm pf}$&$\Sigma_+$&$\Sigma_-$&$\Sigma_\times$&$N_-$&$\Theta$&$V_1$\\
\hline
$\mathcal{C}^0$(II)& CS & $-1+\frac{3}{2}\gamma$& $(\frac{2}{3},2)$&$1$&0&$\frac{3}{16}(6-\gamma)$&$-\frac{3}{16}(\gamma-\frac{2}{3})$&$\pm\sqrt{3}\frac{3}{16}(\gamma-\frac{2}{3})$&0&$\pm\frac{3}{8}\sqrt{(2-\gamma)(\gamma-\frac{2}{3})}$&0&0\\
$\mathcal{C}^+$(VI$_h$)&	PW$(\alpha,\beta_1,\nu^2)$& $-2\beta_1$& $[0,2]$&$0<\alpha^2<1$&$1+\beta_1$&0&$\beta_1\,\leq\,0$&0&$-\nu$& $\nu$&$-V_1$&$\pm\sqrt{-\beta_1(1+\beta_1)-\nu^2}$ \\
{\color{gray}$\mathcal{S}^0$(V)}&{\color{gray}M}&{\color{gray} 0}& {\color{gray}$[0,2]$}&{\color{gray}free}&{\color{gray}$1$}&{\color{gray}0}&{\color{gray}0}&{\color{gray}0}&{\color{gray}0}&{\color{gray}0}&{\color{gray}0}&{\color{gray}0}\\
\hline
	\end{tabular}}
	\caption{Summary of equilibrium sets $\mathcal{P}$ analyzed in $N_-$\,-\,gauge, where $N_+=\sqrt{3}\alpha N_-$. Here $\beta_1\,>\,-1$. M is per definition part of PW($\alpha, \beta_1,\nu^2$), and is therefore shadow-faced.} 
	\label{tab:VIhNm}
\end{table}
\begin{table}[H]
	\centering
	\resizebox{\textwidth}{!}{\begin{tabular}{llcccccccccc}
			\toprule
			\multicolumn{12}{c}{\textbf{Equilibrium sets of $\overline{\mathcal{B}(\textrm{VI}_{\tilde{h}}})$ analysed in $F$\,-\,gauge.}} \\
			\hline
Set&$\mathcal{P}$ & $q$ & $\gamma $&$A$&$\Omega_{\rm pf}$&$\Sigma_+$&$\Sigma_-$&$\Sigma_\times$&$\nu^2$&$\Theta$&$V_1$\\
\hline
$\mathcal{S}^0$(I)& flat FLRW & $-1+\frac{3}{2}\gamma$& $[0,2]$&$0$&1&0&0&0&0&0&0 \\
$\mathcal{S}^0$(V)& open FLRW & 0& $\frac{2}{3}$&$A\,\in\,[0,1]$&$1-A^2$&0&0&0&0&0&0\\
$\mathcal{C}^0$(I)&JED$(\beta_1,\beta_2,\beta_3)$&2&$[0,2)$&0&0&$\beta_1$&$\beta_2$&$\beta_3$&0&$[-\sqrt{1-\beta^2},\sqrt{1-\beta^2}]$&0\\
{\color{gray}$\mathcal{C}^0$(I)}&{\color{gray} K$(\beta_1,\beta_2)$}&{\color{gray}2}&{\color{gray}$[0,2)$}&{\color{gray}0}&{\color{gray}0}&{\color{gray}$\beta_1$}&{\color{gray}$\beta_2$}&{\color{gray}$[-\sqrt{1-\beta_1^2-\beta_2^2},\sqrt{1-\beta_1^2-\beta_2^2}]$}&{\color{gray}0}&{\color{gray}0}&{\color{gray}0}\\
$\mathcal{C}^0$(I)&JS$(\beta_1,\beta_2,\beta_3,\Theta)$&2&$2$&0&$\sqrt{1-\beta^2-\Theta^2}$&$\beta_1$&$\beta_2$&$\beta_3$&0&$\Theta$&0\\
$\mathcal{S}^+$(VI$_h$)&	W($\kappa,\nu^2$)& $-1+\frac{3}{2}\gamma$& $(\frac{2}{3},2)$&$-\frac{3}{4}(2-\gamma)\kappa$&$\frac{3}{4}(2-\gamma)(1-\kappa^2)$&$ \frac{1}{2} -\frac{3}{4}\gamma$&-$\kappa\nu_3$&$\kappa\nu_2$&$\nu^2$& $\kappa\,V_1$&$\mp\frac{3}{4} \sqrt{(2-\gamma ) (\gamma-\frac{2}{3})-\frac{16}{9}\nu^2}$ \\
{\color{gray}$\mathcal{S}^0$(VI$_h$)}&	{\color{gray}C($\kappa$)}& {\color{gray}$-1+\frac{3}{2}\gamma$}& {\color{gray}$(\frac{2}{3},2)$}&{\color{gray}$-\frac{3}{4}(2-\gamma)\kappa$}&{\color{gray}$\frac{3}{4}(2-\gamma)(1-\kappa^2)$}&{\color{gray}$ \frac{1}{2} -\frac{3}{4}\gamma$}&{\color{gray}-$\kappa\nu_3$}&{\color{gray}$\kappa\nu_2$}&{\color{gray}$\frac{9}{16}(2-\gamma)\left(\gamma-\frac{2}{3}\right)$}&{\color{gray} 0}&{\color{gray}0}\\
\hline
	\end{tabular}}
	\caption{Summary of equilibrium sets $\mathcal{P}$ analyzed in $F$\,-\,gauge. For brevity, notation is such that $\beta^2\,\equiv\,\beta_1^2+\beta_2^2+\beta_3^2$. The parameter $\kappa$ is restricted according to $-1<\,\kappa\,\leq 0$. The group parameter $\tilde{h}$ is negative. K and C are per definition part of JED$(\beta_1,\beta_2,\beta_3)$ and W$(\kappa,\nu^2$), respectively, and are therefore shadow-faced.}
\label{tab:VIhGg}
\end{table}

\begin{figure}[h]
    \centering
    \begin{minipage}{0.4\textwidth}
        \centering
        \includegraphics[width=1\textwidth]{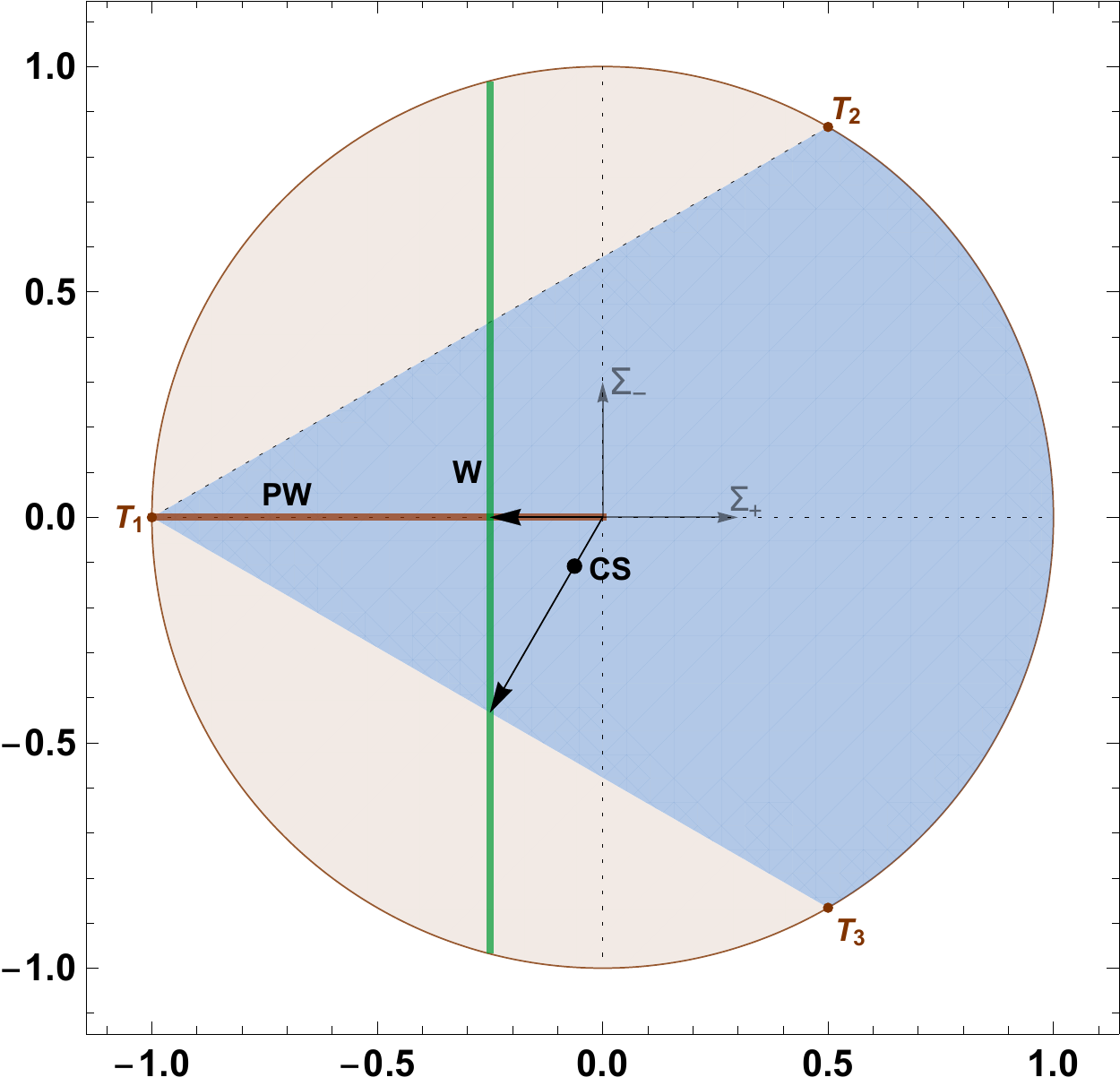} 
    \end{minipage}\hfill
    \begin{minipage}{0.6\textwidth}
        \centering
        \includegraphics[width=1\textwidth]{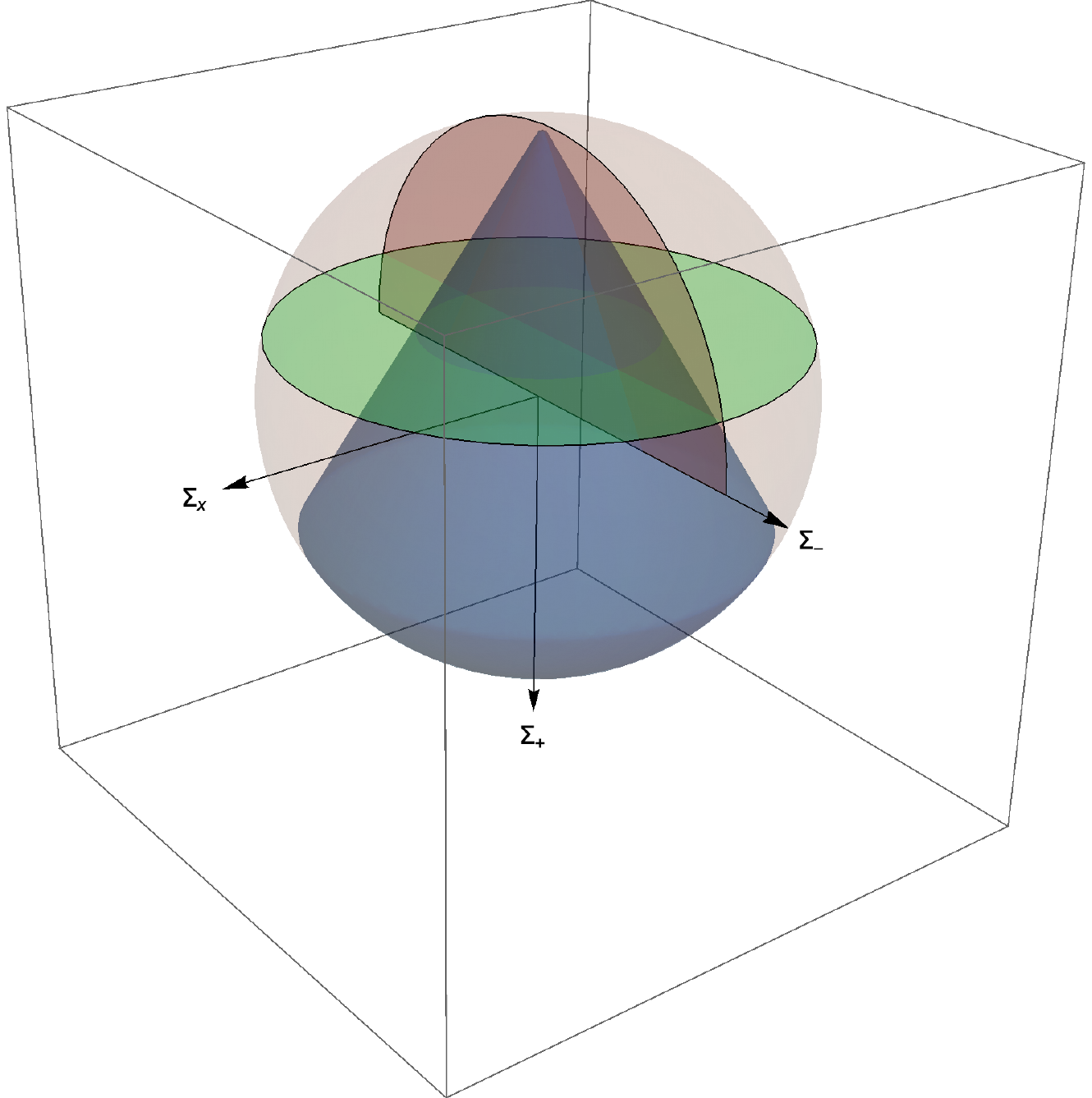} 
        \caption{Here JED (blue and brown-gray) is viewed as a disk in two dimensions (left) and as a sphere in three dimensions $(\Sigma_+,\Sigma_-,\Sigma_\times)$. Projections of W (green), PW (red) and CS (black) are shown. The blue region is the past stable part of JED. The arrows show the directions along which CS and W move as functions of increasing $\gamma$.}
    \end{minipage}
\end{figure}

\subsection{Discussion of stability}
In a previous work \cite{normann20b} we analysed $\mathcal{B}(\textrm{VII}_h)$, and did so without specifying the sign of the group parameter $h$. The analysis revealed that the real parts of all the eigenvalues were independent of $h$ (refer to Tables C1 and C2 in the paper). As a result the treatment in our previous work suffices to establish the local stability also in the present case, where $h<0$. This is true for all equilibrium sets that also exist for these values of $h$. The only equilibrium set not analysed in the previous work is $W(\kappa,\nu^2)$, which is the part of the Wonderland fabric contained in $\mathcal{B}$(VI$_{\tilde{h}}$). This set will therefore receive special attention in this section. Table \ref{tab:StabVIh} summarizes the results. 

\subsubsection{Wonderland W($\kappa,\nu^2$). } The type $\mathcal{B}$(VI$_{\tilde{h}}$)\setcounter{footnote}{0}\footnote{Where, remember, $\tilde{h}\,\leq\,0$ and $\tilde{h}\,\neq\,-1\,\cup\,-1/9$.} subset of Wonderland is W$(\kappa,\nu^2)\,\equiv\,\lim_{\nu_1\to 0}$W$(\kappa,\nu_1,\nu^2)$. Refer to Table \ref{tab:VIhGg} for specification of $q$ and $\Omega_{\rm pf}$ in this set. 

Performing the local stability analysis, we find that in the 8-dimensional physical state space, three zero-eigenvalues correspond to perturbations into the set itself (the parameters $\kappa,\nu_2,\nu_3$)\footnote{If the reader is puzzled by the number of zero-eigenvalues: There must necessarily be one zero-eigenvalue for each parameter in the equilibrium set. This is so, since no dynamics happen in the set. Any zero-eigenvalues surplus of the number of parameters would however render the first-order analysis inconclusive, but this is not the case here.}. The five remaining eigenvalues are always negative, and read as follows.

\begin{eqnarray}\fl
\label{Weigenv}
\left\{-\frac{3}{2}(2-\gamma),-\frac{3}{4} (2-\gamma ) \left(1\,\pm\,\sqrt{B(\gamma,\kappa)}\right),-\frac{3}{4} (2-\gamma )\,\pm\,\frac{\sqrt{3}}{4}\sqrt{C(\gamma,\kappa,\nu^2)}\right\}.
\end{eqnarray}
Here
\begin{equation}\fl
\label{B}
B(\gamma,\kappa)=6 \gamma  \left(\kappa ^2-1\right)-4 \kappa ^2+5
\end{equation}
and
\begin{equation}\fl
\label{C}
C(\gamma,\kappa,\nu^2)=3 (\gamma -2)^2+64 \left(\kappa ^2-1\right) \nu ^2.
\end{equation}
Since the eigenvalues are negative, W$(\kappa,\nu^2)$ is stable, and hence Wonderland is (at least locally) a future attractor.

\paragraph{The Collins solution. } It is noteworthy that the Wonderland solution reduces to the Collins solution in the absence of the $j$-form fluid. Putting $V_1=\Theta=0$ in the specifications of Wonderland above, one finds the further specialisation
\begin{eqnarray}
\label{nu}
\nu^2=\frac{9}{16}(2-\gamma)\left(\gamma-\frac{2}{3}\right).
\end{eqnarray}
Together with \eref{Wspec1}-\eref{Wspec2} this gives the Collins self-similar solution. From the specifications of Wonderland one finds $A^2=9\kappa^2(2-\gamma)^2/16=-\tilde{h}\nu^2$. Hence, with eq. \eref{nu} one has that 
\begin{equation}
\label{h2}
    \tilde{h}=-\frac{2-\gamma}{3\gamma-2}\kappa^2.
\end{equation}
From the expression for $\Omega_{\rm pf}$ in Wonderland one finds that $-1\,<\,\kappa\,\leq\,0$. Hence, from Eq. \eref{h2} it is evident that $\tilde{h}\,>\,-(2-\gamma)/(3\gamma-2)$. This result agrees with previous treatments, for instance \cite[Sec.~7.2.2.]{bok:EllisWainwright}.
\paragraph{The other equilibrium sets: } For $\gamma\,\in\,(0,2/3)$ we have already seen that a previously established theorem ensures that FLRW is the global future attractor. We also see from Table \ref{tab:StabVIh} that Jacobs' Extended Disk (JED) and Jacobs' Sphere (JS) are the only options if there exist past global attractors. Note that the Kasner (K) space-time is a subset of JED. The point $\gamma=2/3$ remains uncertain, although somewhat constrained (we must have $1=\Omega_{\rm pf}+A^2$ by~\cite[Prop. 2]{normann20a}, as mentioned earlier). The Plane Waves (PW) include the Milne vacuum solution, and exists for all values of $\gamma$. In the absence of the perfect fluid it is, as discussed earlier, an attractor whereas in the presence of the perfect fluid its stability changes, as shown in Table \ref{tab:StabVIh}. The Collins-Stewart solution (CS) is also present, but with two positive eigenvalues it is a saddle.

\paragraph{}Table \ref{tab:StabVIh} summarizes the overall stability of the equilibrium sets found in $\mathcal{B}$(VI$_h$).
\begin{table}[H]
	\centering
	\resizebox{\textwidth}{!}{\begin{tabular}{lcclclclclc}
			\toprule
			\multicolumn{11}{c}{\textbf{Classification of equilibrium sets in $\overline{\mathcal{B}(\textrm{VI}_{\tilde{h}}})$}} \\
			\hline
$\mathcal{P}$&& Existence && Attractor && Saddle && Repeller && Inconclusive \\
			\hline
PW($\alpha,\beta,\nu^2$) && $\gamma\in[0,2]$ &&$\beta_1\,>\,-\frac{3}{4}\left( \gamma-\frac{2}{3}\right)$&&$\beta_1\,<\,-\frac{3}{4}\left( \gamma-\frac{2}{3}\right)$&&&&$\beta_1\,=\,-\frac{3}{4}\left( \gamma-\frac{2}{3}\right)$\\
W($\kappa,\nu^2$)&& $\gamma\in(\frac{2}{3},2)$&& $\forall\, \kappa,\nu,\gamma$  && &&&& \\
open FLRW  && $\gamma = \frac{2}{3}$ &&&&&&&&$\forall$\\
flat FLRW && $\gamma\in[0,2)$ &&$\gamma\,\in\,[0,\frac{2}{3})$&&$\gamma\,\in\,(\frac{2}{3},2)$&&&&$\gamma=\frac{2}{3}$\\
K$(\beta_1,\beta_2)$&& $\gamma\in[0,2)$ &&&&\textrm{else}&&$\beta_1>\frac{1}{2}$&&$\beta_1=-1\,\cup\,\frac{1}{2}$\\
JED$(\beta_1,\beta_2,\beta_3)$&& $\gamma\in[0,2)$ &&&&\textrm{else}&&$\beta_1>-1+\sqrt{3}\sqrt{\beta_2^2+\beta_3^2}$&&$\beta_1=-1+\sqrt{3}\sqrt{\beta_2^2+\beta_3^2}$\\
JS$(\beta_1,\beta_2,\beta_3,\Theta)$&& $\gamma=2$ &&&&\textrm{else}&&$\beta_1>-1+\sqrt{3}\sqrt{\beta_2^2+\beta_3^2}$&&$\beta_1=-1+\sqrt{3}\sqrt{\beta_2^2+\beta_3^2}$\\
CS&&$(\frac{2}{3},2)$&&&&$\forall\,\gamma$&&&&\\\hline
	\end{tabular}}
	\caption{The domains where the local stability analysis is conclusive are divided into attractor, saddle and repeller subdomains.  The rightmost column shows the domains where the linear stability analysis is inconclusive.}
	\label{tab:StabVIh}
\end{table} 

\section{The invariant set $\mathcal{B}$(VI$_0$)}
\label{Sec:VI0}
The closure of $\mathcal{B}$(VI$_0$) is
\begin{equation}
\overline{\mathcal{B}(\textrm{VI}_0)}=\mathcal{B}(\textrm{VI}_0)\cup\mathcal{C}(\textrm{II})\cup\mathcal{C}(\textrm{I}).
\end{equation}
With $A=0$ it is evident from the equation for $\Theta'$ (refer to~\ref{App:dynsys}) that the time-like part of the $j$-form field will vanish asymptotically (except for $q=2$). The spatial part of the form-field, however, will, as we shall see, generally not die away for $\gamma\,>\,2/3$. The key to analysing this 7-dimensional dynamical system, is the monotonic function
\begin{eqnarray}
\label{b1}Z_6=\frac{V_1^{3\gamma-2}\Omega^2}{\phi^{3\gamma+2}}, \qquad \phi=1+m\Sigma_+, \qquad m=\frac 14(3\gamma-2), \\
\label{b2}\frac{Z_6'}{Z_6}=\phi^{-1}\left[8(\Sigma_++m)^2+\frac 32(3\gamma+2)(2-\gamma)\left(|{\mathbf\Sigma}_{\Delta}|^2+\Theta^2\right)\right].
\end{eqnarray} 
This monotonic function was also used in $\mathcal{B}$(VII$_0$) (cf.~\cite{normann20a}) which is the same dynamical system, modulo the type constraint~\ref{grConstr}, which has the opposite sign. The consequence of the difference in group constraint leads to the existence of a different future attractor: Wonderland in type VI$_0$; $W(\nu^2)\,=\,\lim_{\kappa\to 0}W(\kappa,\nu^2)\,\subset\,\mathcal{B}\textrm{(VI}_{\tilde{h}}$). On the other hand, the resemblance to $\mathcal{B}$(VI$_0$) described above, allows us to use the results obtained for $\mathcal{B}$(VI$_0$), the only difference being that the future attractor is now  $W(\nu^2)$ rather than $W(\nu_1).$

This conclusion is supported by the eigenvalues given in Eq. \eref{Weigenv}. For $\kappa=0$ we find that the eigenvalues are still all eigenvalues negative, except if (i) $(\nu_2,\nu_3)\,\to\,(0,0)$, in which case there will be two extra zero eigenvalues, (ii) or if $\gamma=6/5$, in which case there will also be one extra zero-eigenvalue. The monotonic function $Z_6$, however, comes to our rescue, and reveals that indeed $W(\nu^2)$ is the global future attractor in the set $\mathcal{B}$(VI$_0$). To summarize, we reach the following global conclusions. JED (JS) is the global past attractor for $\gamma<2$ ($\gamma=2$). For $\gamma\,\leq\,2/3$ FLRW is the global future attractor. $W(\nu^2)$ is the global future attractor in the interval $2/3\,<\gamma\,<2$, and the point $\Sigma_+=-1$, where JED, JS and Wonderland meet is the the global future attrator for $\gamma=2$. We refer the reader to our previous paper for further details. 
\begin{figure}[H]
\centering
\begin{tikzpicture}
\node[inner sep=0pt] (russell) at (0,0)
    {
    \begin{overpic}[width=0.7\textwidth,tics=10]{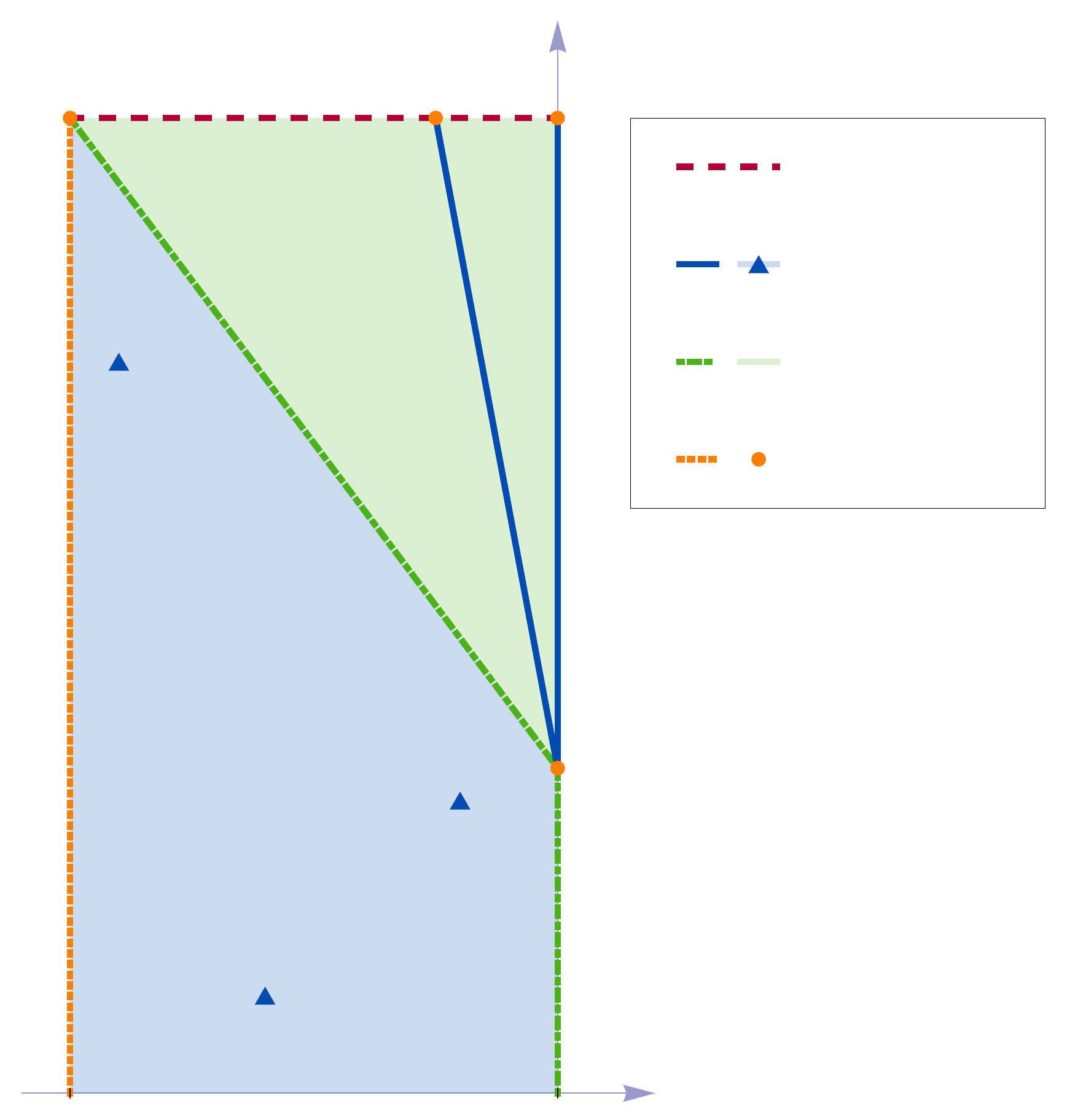}
\put (47,10) {\scriptsize{\rotatebox{90}{flat FLRW}} }
\put (47,65) {\scriptsize{\rotatebox{90}{flat FLRW}} }
\put (13,82) {\scriptsize{\rotatebox{-53.5}{Wonderland (Collins when $\Theta=V_1=0$)}} }
\put (17,25) {\scriptsize{\rotatebox{36.5}{Plane Waves}} }
\put (25,75) {\scriptsize{\rotatebox{36.5}{Plane Waves}} }
\put (40.6,65) {\scriptsize{\rotatebox{-78.3}{Collins-Stewart}} }
\put (25,90.5) {\scriptsize Jacobs' Sphere}
\put (7.3,50) {\scriptsize{\rotatebox{-90}{Kasner}} }
\put (50,96) { \transparent{0.5}%
{\textcolor{axisColor}{$\displaystyle \gamma$ }}}
\put (3.4,0) { \scriptsize $\displaystyle -1$ }
\put (51.7,31) {\scriptsize \transparent{0.5}%
{\textcolor{axisColor}{$\displaystyle \frac{2}{3}$ }}}
\put (51.7,88.7) {\scriptsize \transparent{0.5}%
{\textcolor{axisColor}{$\displaystyle 2$ }}}
\put (58,1.6) { \transparent{0.5}%
{\textcolor{axisColor}{\scriptsize $\displaystyle \Sigma_+$ }}}	

\put (76,84.5) {\scriptsize{Repeller}}
\put (76,76) {\scriptsize{Saddle}}
\put (64.5,76) {\scriptsize{$/$}}
\put (76,66.9) {\scriptsize{Attractor}}
\put (64.5,66.9) {\scriptsize{$/$}}
\put (76,58.5) {\scriptsize{Inconclusive}}

	\end{overpic}};
\node[] at (-0.31,-5.8) {\scriptsize $\displaystyle \Omega_{\rm pf}=1\,\uparrow$};
\node [align=left] at (3.1,-2.3) {
Schematic illustration \\
showing how the\\
different attractors \\
(green) relate to each \\
other. Note that $\gamma$ \\
increases upwards, and \\
$\Sigma_+$ is non-positive.\\
Also, Collins-Stewart\\
and FLRW are in the \\
boundaries.
};
\end{tikzpicture}
\caption{$\overline{\mathcal{B}\textrm{(VI}_{\tilde{h}})}$.}
\label{fig:B6h}
\end{figure}
\newpage
\section{A closer look at the anisotropic, self-similar assymptotic states}
\label{ref:Anis}
In this section we take a closer look at the self-similar line-elements corresponding to the anisotropic equilibrium sets; Wonderland in particular. The fact that the scalar variables of the theory take constant values at the equilibrium sets (in expansion normalized variables) makes it possible to write down the line-elements corresponding to exact, self-similar solutions to Einstein's theory of gravity. In particular, the shear-tensor $\Sigma_{ij}$, the isotropic expansion $H$ and the deceleration parameter $q$ are observables that are related to the geometry of the spatial hypersurfaces through the equation
\begin{equation}
    \frac{1}{2}\mathcal{L}_{\partial_t}\mathbf{h}=\bm{\theta}.
\end{equation}
Here $\mathcal{L}$ denotes the Lie derivative, $\mathbf{h}$ is the spatial metric and $\bm{\theta}=H(\delta_{ij}+\Sigma_{ij})\omega^i\otimes\omega^j$\setcounter{footnote}{0}\footnote{not to be confused with $\Theta$, which is the expansion-normalized time-like component of the $j$-form fluid, cf.~\eref{J}.} is the expansion tensor. Since we use orthonormal frames we have $\mathbf{h}=\delta_{ij}\omega^i\otimes\omega^j$, where $\{\omega^i\}$ are basis 1-forms. Starting from a general, self-similar line-element of Bianchi type VI, one may obtain the line-element corresponding to each equilibrium point. The self-similar cosmological models obtained as equilibrium sets in the dynamical systems of the orthonormal-frame approach, are on the form~\cite{rosquist88}
\begin{eqnarray}
\label{lineEl}
 {\rm d}s^2 &=-{\rm d}t^2+t^{2a}\biggl[t^{4p}\left(W^1+bt^cW^2+rt^sW^3\right)^2 \\
 &\phantom{=}+t^{-2p+2w}\left(W^2+ut^vW^3\right)^2+t^{-2p-2w}(W^3)^2 \biggr],
\end{eqnarray}
where $\{W^i\}$ are left-invariant one-forms that fulfill the Lie algebra associated with a certain Bianchi type. In our calculations we have used an orthonormal frame, and the expansion-tensor $\bm{\theta}$ of the equilibrium set is henceforth given in that frame. We must therefore relate the general form of the line-element above to the orthonormal frame. To this end we find an orthonormal basis:
\begin{eqnarray}
\label{omega}
&\omega^0={\rm d}t,\\
&\omega^1=t^{a+2p}(W^1+bt^cW^2+rt^sW^3),\\
&\omega^2=t^{a-p+w}(W^2+ut^vW^3),\\
&\omega^3=t^{a-p-w}W^3.
\end{eqnarray}
In this basis the metric $\mathbf{g}=g_{\mu\nu}\omega^\mu\otimes\omega^\nu\,$ simplifies to $\mathbf{g}\,=\,\eta_{\mu\nu}\omega^\mu\otimes\omega^\nu$, where $\eta_{\mu\nu}$ is the Minkowski metric. As we shall see in a later section, these equations will be important also when we want to go the other way: writing down the general expressions for the gauge potential in a coordinate basis, starting from expressions in the orthonormal frame.

As a final point we might also want to write down the equation-of-state parameter $\xi$ of the $j$-form matter of the various equilibrium sets. It is given by~\cite{normann18}
\begin{equation}
    \xi=\frac{\Theta^2-(V_1^2+\abs{\mathbf{V}_c}^2)/3}{\Theta^2+V_1^2+\abs{\mathbf{V}_c}^2}+1,
\end{equation}
and in our case $\mathbf{V}_c=0$ throughout. In the following, we look closer at the equilibrium sets W, PW and JED/JS.

\subsection{Wonderland}
For Wonderland, where $\Theta^2=\kappa^2\,V_1^2$, this parameter becomes
\begin{equation}
    \xi_{\,_{\rm W}}=2\frac{\kappa^2+1/3}{\kappa^2+1}.
\end{equation}
Hence the e.o.s.-parameter for Wonderland lies in the interval 
\begin{equation}
    \frac{2}{3}\,\leq\,\xi_{\,_{\rm W}}\,\leq\,\frac{4}{3}.
\end{equation}
Since $\kappa$ is a parameter that does not change, this means that $\xi_{\rm W}$ does not change with time. The line-element that corresponds the Wonderland equilibrium set of Bianchi-type VI$_{\tilde{h}}$ is 
\begin{equation}
\fl{\rm d}s^2=-{\rm d}t^2+t^2{\rm d}x^2+t^{\frac{2-\gamma}{\gamma}-\Gamma}\left(e^{-kfx}{\rm d}y+\frac{\nu_2}{\nu_3} t^{\Gamma}e^{-kx}{\rm d}z\right)^2+t^{\frac{2-\gamma}{\gamma}-\Gamma}e^{-2kx}{\rm d}z^2
\end{equation}
where $\Gamma$ is actually a function (arguments suppressed for brevity of notation) such that $\Gamma(\nu_3,\gamma)=4\kappa\nu_3/(\sqrt{3}\gamma)$. Also, $-1\,\leq\,f\,<\,1$. Since it is $\nu^2=\nu_2^2+\nu_3^2$ that is the scalar, and not $\nu_2$ and $\nu_3$ themselves, one may wonder how they behave. As shown in \cite{normann20a}, however, the evolution of $\nu_2$ and $\nu_3$ is given by
\begin{eqnarray}
\label{nu2d}
\nu_2'=-2\kappa\nu_1\nu_3,\\
\label{nu3d}\nu_3'=2\kappa\nu_1\nu_2.
\end{eqnarray}
Hence, since $\nu_1=0$ in this case, one finds $\nu_2'=\nu_3'=0$. 
\subsection{Plane Waves and JED/JS}
A similar calculation for the Plane Waves solution, where $\Theta^2=V_1^2$, shows that
\begin{equation}
    \xi_{\,_{\rm PW}}=\frac{4}{3}.
\end{equation}
The line-element of this solution may be shown to be
\begin{equation}
\label{metricPW}
{\rm d}s^2=-{\rm d}t^2+t^{2}{\rm d}x^2+t^{-1+\frac{3}{1-2\beta_1}}\left(\left(e^{-kfx}{\rm d}y+ue^{-kx}{\rm d}z\right)^2+e^{-2kx}{\rm d}z^2\right),
\end{equation}
where $-1<\Sigma_+=\beta_1<0$. For the past attractors, JED and JS, one has $V_1=0$, and hence $\xi_{\,_{\rm J}}=2$. The line-element of these solutions is
\begin{equation}
\label{metricJED}
{\rm d}s^2=-{\rm d}t^2+t^{\frac{2}{3}(1-2\beta_1)}{\rm d}x^2+t^{\frac{2}{3}(1+\beta_2+\sqrt{3}\beta_3)}({\rm d}y+ut^{\frac{2}{\sqrt{3}\beta_3}}{\rm d}z)^2+t^{\frac{2}{3}(1+\beta_2-\sqrt{3}\beta_3}{\rm d}z^2.
\end{equation}
which, with $u=0$ reduces to the Kasner solution (e.g. cf. Sec. 13.2 in \cite{bok:GronHervik}).
JED and JS have unstable regions, and hence, in the case where there is no closed orbit, one must have 
\begin{equation}
    2=\lim_{t\to-\infty}\xi\,>\,\lim_{t\to\infty}\xi\,\in\,[2/3,4/3]=\xi_{\rm W}.
\end{equation}
We note that $\xi=1$ (corresponding to dust) is in this interval. 
\begin{figure}[h]
	\centering	
	\begin{overpic}[width=0.8\textwidth,tics=10]{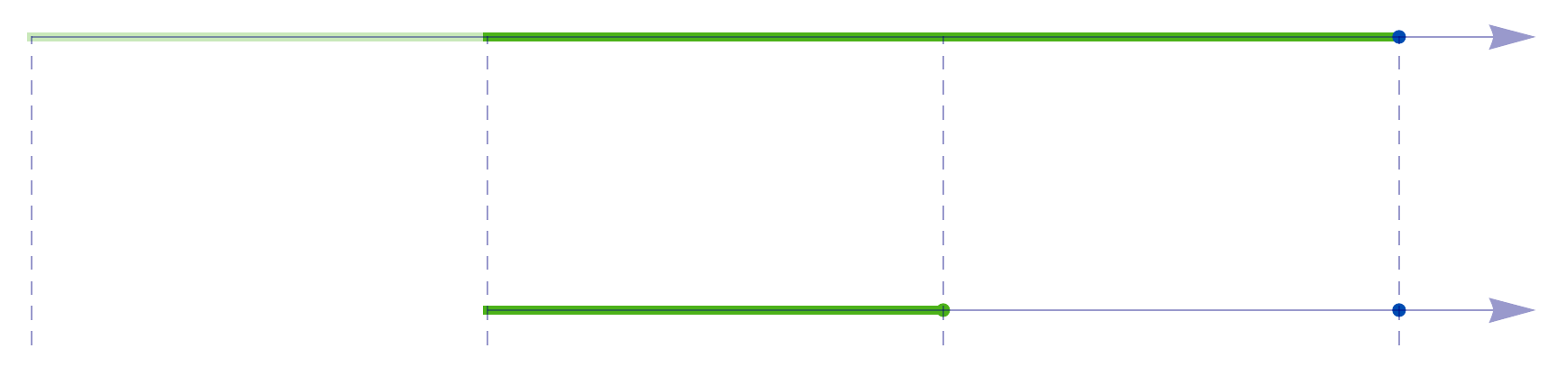}	
		\put (100,21) { $\displaystyle \gamma$}	
		\put (100,3.1) { $\displaystyle \xi$ }	
		\put (0.3,-0.5) { \scriptsize $\displaystyle 0$ }
		\put (29.35,-0.5) {\scriptsize $\displaystyle 2/3$ }
		\put (57.45,-0.5) { \scriptsize $\displaystyle 4/3$ }
		\put (87.64,-0.5) { \scriptsize $\displaystyle 2$ }
	\end{overpic}	
	\caption{The diagram shows the parameter regions for which future attractor solutions exist (green). Light green is FLRW and hence isotropic. Solid green is W and PW. $\xi$ is defined only for values $\,\in\,[2/3,2]$ and future attractors are found only for $\xi\,\in\,[2/3,4/3]$. The past attractors (blue dots) are JED and JS and are confined to extremal parameter values.} 
	\label{Fig:stability_BVII0}
\end{figure}
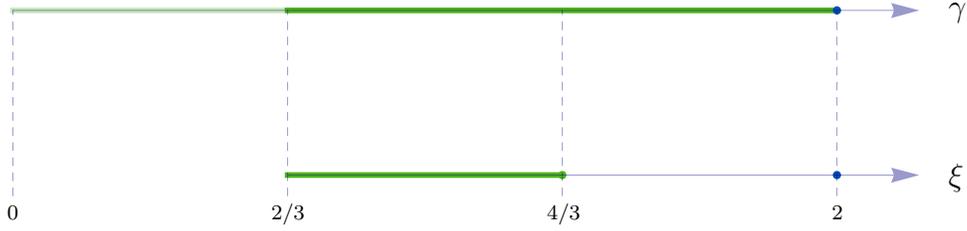

\section{A Closer look at Wonderland}
\subsection{Realization of Wonderland by mass-less scalar field}
\label{Sec:scalarfield}
As mentioned, the current study may be viewed as that of a inhomogeneous, mass-less scalar gauge field. Taking $j=1$ one could take the $j$-form $\mathcal{J}$ to be constructed from a mass-less scalar gauge-field $\phi=\phi(t,\mathbf{x})$, which is a $0$-form. Id est; let
\begin{equation}
\label{relapfi}
\ext{\phi}=\mathcal{J}\quad\rightarrow\quad\mathcal{J}_\mu=\partial_\mu\phi.
\end{equation}
As before, we decompose the $1$-form according to
\begin{equation}
\label{decomp}
\mathcal{J}_\alpha=-w\, u_\alpha+v_\alpha\,,
\end{equation}
where the 4-velocity $u_\alpha$ is time-like ($u_\alpha u^\alpha\,<\,0$), whereas $v_\alpha$ is defined to be orthogonal to $u_\alpha$ and therefore space-like ($v_\alpha v^\alpha\,>\,0$). Note next that according to the decomposition in \ref{App:Decomp} we have 
\begin{equation}
\label{vtheta}
  v_i=\sqrt{6}H V_i\quad\quad\quad\textrm{and}\quad\quad\quad w=\sqrt{6}\Theta H.
\end{equation}
From the implicit definition of $q$ according to $\dot{H}=-(1+q)H^2$ we find the relation
\begin{equation}
    H=\frac{1}{(1+q)t}.
\end{equation}
\textbf{Word of caution:} It is important in the following, that we recall the necessary transition from the orthonormal frame to a coordinate basis $\{t,x,y,z\}$ through the relations \eref{omega}. More specifically; we intend to calculate equation~\eref{relapfi} in a coordinate basis. We have previously, however, calculated $\mathcal{J}_\mu$ in an expansion-normalized orthonormal frame. Denoting the orthonormal frame by $\hat{\,}$ we find
\begin{equation}
\partial_\mu\phi=\frac{\partial x^{\hat{\nu}}}{\partial x^\mu}\partial_{\hat{\nu}} \phi=\frac{\partial x^{\hat{\nu}}}{\partial x^\mu}\mathcal{J}_{\hat{\nu}}.
\end{equation}
The transformation rules $\frac{\partial x^{\hat{\nu}}}{\partial x^\mu}$ must now be calculated from \eref{omega} for each individual case.

\paragraph{Considering Wonderland.} For the equilibrium set Wonderland we have
\begin{equation}
    V_2=V_3=0=\Sigma_2=\Sigma_3=\Sigma_-=0.
\end{equation}
Now; $V_2=V_3=0$ makes it clear that it is only $V_1$ we need to transform to a coordinate basis. Second; $\Sigma_2=\Sigma_3=0$ gives $b=r=0$. Hence, using also that $W^1={\rm d}x$ in all the Bianchi types we find from~\eref{omega} that
\begin{equation}
\frac{\partial x^{\hat{1}}}{\partial x^x}=\frac{\partial \omega^1}{\partial x^x}=t^{a+2p}
\end{equation}
in all the invariant sets above. Furthermore, since $q$ and $\Sigma_+$ are the same in all of the Wonderland fabric, we find, in all the invariant sets, that
\begin{equation}
    a+2p=1.
\end{equation}
Hence, for the Wonderland fabric we have (using also \eref{vtheta}) that
\begin{equation}
    \partial_x\phi=\frac{2\sqrt{6}}{3\gamma}V_1,\quad\quad \partial_t\phi=\frac{2\sqrt{6}}{3\gamma t}\Theta.
\end{equation}
Integrating out, we find
\begin{equation}
\label{sol}
\fl\phi(x,t)=C(\gamma)\left(x+\kappa\ln{t}\right)\quad\quad\textrm{where}\quad\quad
C(\gamma)=\frac{2\sqrt{6}}{3\gamma}\frac{3}{4}\sqrt{(\gamma-\frac{2}{3})(2-\gamma)-\frac{16}{9}\nu^2}.\end{equation}
The particular solutions will hence look different for the different invariant Bianchi-sets. For instance; $\nu^2=0$ in all the sets except Type VI. Also, in Type I $\Theta=0$ (id est; $\kappa=0$), and the time-dependence therefore vanishes. The energy density is given by~\cite{normann18}
\begin{equation}
    \rho=\frac{1}{2}\left(v_1^2+w^2\right).
\end{equation}
Inserting from above one finds that this gives 
\begin{equation}
\rho(\gamma,t)=\left(\frac{C(\gamma)}{3t}\right)^2\left(1+\kappa^2\right),
\end{equation}
where $C(\gamma)$ is specified in Eq.~\eref{sol}. We therefore find that the energy density decreases as $\rho\,\sim\,t^{-2}$. As a final note to the reader puzzled by the fact that $\phi$ increases with $x$ and $t$: Note that it is the field-strength $\mathcal{J}$ that represent the physical variables. Equivalently; only \textit{changes} in $\phi$ can possibly be physical, since the equations of motion remain unchanged under a change $\phi\rightarrow\phi+{\rm k}$ for any constant $k$. This is merely a reflection of the fact that we have constructed a gauge theory.
\subsection{Simulations}
For illustrational purposes we also provide some simulations, although we restrict the analysis to $\mathcal{B}$(I) for the sake of simplicity. Also, the spatial background geometry of this Bianchi type is Euclidean.

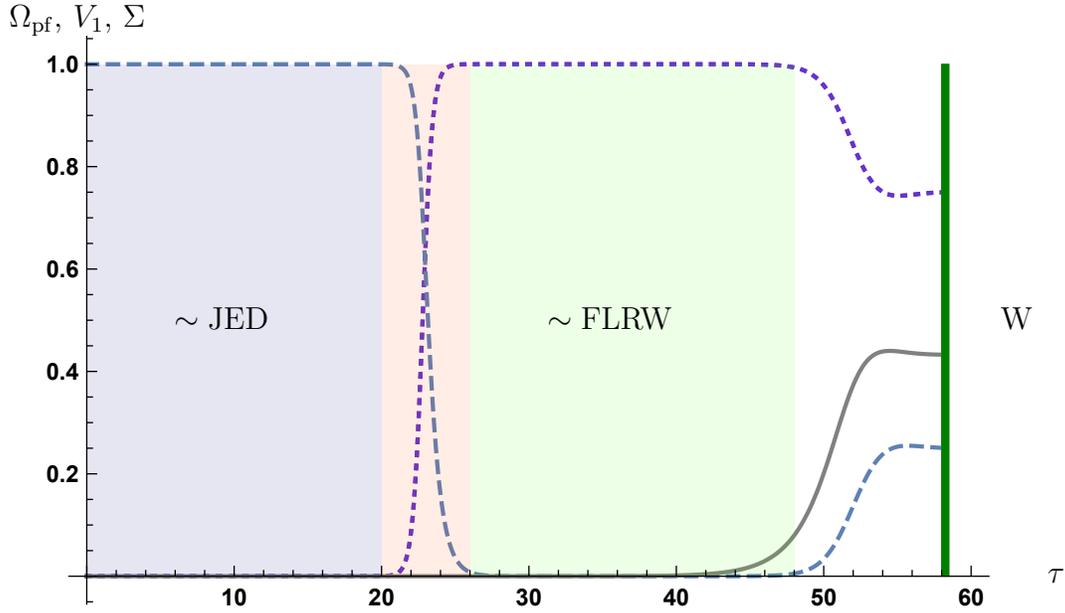
\begin{figure}[h]
	\centering	
	\begin{overpic}[width=0.8\textwidth,tics=10]{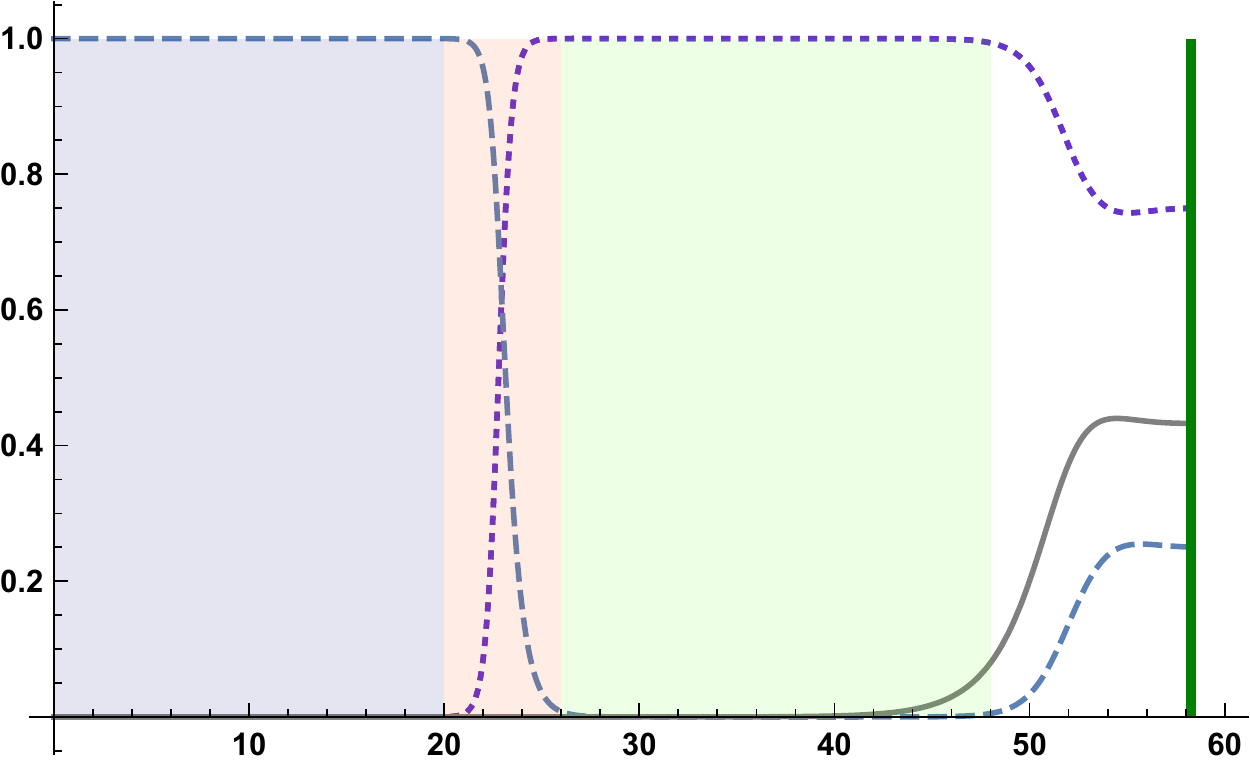}
		\put (-5,62) { $\displaystyle \Omega_{\rm pf},\,V_1,\,\Sigma$}
		\put (100,30) { $\displaystyle \textrm{W}$ }
		\put (12.5,30) { $\displaystyle \sim\textrm{JED}$ }
		\put (52,30) { $\displaystyle \sim\textrm{FLRW}$ }
		\put (105,3.1) { $\displaystyle \tau$ }	
	\end{overpic}	
	\caption{Simulation showing the development of the shear $\Sigma$ (blue, dashed), the perfect-fluid energy density $\Omega_{\rm pf}$ (purple, dotted) and the $j$-form component $V_1$ (grey) for $\gamma=1$ in $\mathcal{B}$(I). On the horizontal axis is dynamical time $\tau$. Note that there will be an JED-like epoch and an FLRW-like epoch before the universe reaches Wonderland (W) at about $\tau=58$ (green line). Refer to the main text for further discussion. } 
	\label{Fig:ApproachingW}
\end{figure}
The simulation in Figure~\ref{Fig:ApproachingW} shows a generic scenario in these kinds of models: The universe starts off in a shear-dominated universe (JED), and at some point (depending on initial conditions) it isotropizes, before it enters an FLRW-like stage for a prolonged period of time. In the simulation in Figure~\ref{Fig:ApproachingW}, this period lasts from about $\tau=26$ until $\tau=48$. Hence $\Delta\tau=22$. The dynamical time $\tau$ is related to coordinate time through the scale factor according to
\begin{equation}
\label{ttau}
    \frac{{\rm d}t}{{\rm d}\tau}=\frac{1}{H}.
\end{equation}
Denoting the scale-factor by $a$, we find $H=\dot{a}/a$. Hence, by~\eref{ttau} we find that the scale-factor is given by
\begin{equation}
a(\tau)=a_0{\rm e}^\tau.
\end{equation}
this means that the FLRW period lasts for about 22 e-folds in this case. During this period, the shear is very small. One finds for instance $\Sigma(\tau=40)\,\sim\,10^{-6}$. After the FLRW-epoch, the universe will develop more dynamically before it eventually approaches Wonderland. 
\newpage
\section{Conclusion}
\label{Sec:Concl}
In this paper we have studied a perfect-fluid model with a mass-less scalar gauge-field with an inhomogeneous gradient and shown the existence of an anisotropic future attractor (Wonderland) in $\mathcal{B}$(VI$_0$) and $\mathcal{B}$(VI$_{\tilde{h}}$), where $\tilde{h}=h\,<\,0\,\cap\,h\,\neq\,-1\,\cup\,-1/9$. Wonderland correspond to a family of exact solutions to the Einstein field equations. Our work shows that a generic class of scalar-field models for dark matter, or otherwise, also in these classes of homogeneous cosmologies will lead to anisotropic future states.

Moreover, through an explicit simulation of a dust-filled universe ($\gamma=1$) in the Bianchi I invariant set with the scalar gauge field, we have shown that there will typically be an intermediate quite isotropic phase lasting for $n$ e-folds. In our particular simulation we found $n=25$. Subsequently, the evolution asymptotically approaches the anisotropic attractor named Wonderland. In the simulation, the isotropization-phase between JED and FLRW lasts for about 6 e-folds.

We have also written down the line-element of Wonderland (alongside JED and Plane Waves). The equation-of-state-parameter of Wonderland was generally (independent of Bianchi type) found to lie in the interval $2/3\,\leq\,\xi_{\,_{\rm W}}\,\leq\,4/3$, and the energy-density $\rho$ of the $j$form fluid in the Wonderland-solution was found to decreases as $\rho\,\sim\,1/t^2$.

\section*{Acknowledgements}
We thank Prof. Georgi Dvali and Dr. Mikjel Thorsrud for useful discussions and valuable input. This work was supported through the Research Council of Norway, Toppforsk
grant no. 250367: \emph{Pseudo-Riemannian Geometry and Polynomial Curvature Invariants:
Classification, Characterisation and Applications.}


\appendix
\newpage
\section{Complex variables}
\label{App:CompVar}
In this appendix, we briefly recapitulate how the expansion-normalized variables used in this paper are formed. Refer to \cite{normann18} for further explanations. The expansion-normalization is as follows. 
\begin{eqnarray}
\label{defENC}
&\fl\Sigma_{+}=\frac{\sigma_+}{H}\,,\quad\quad \Omega_i=\frac{\rho_i}{3H^2}\phantom{0},\quad\quad A_i=\frac{a_i}{H}\phantom{000.},\nonumber\\
&\fl\Sigma_{-}=\frac{\sigma_-}{H}\,,\quad\quad \Omega_\Lambda=\frac{\Lambda}{3H^2}\,,\quad\quad N_{+}=\frac{n_+}{H}\phantom{00.},\nonumber\\
&\fl\Sigma_{\times}=\frac{\sigma_\times}{H}\,,\quad\quad V_i=\frac{v_i}{\sqrt{6}H}\,,\quad\quad N_{-}=\frac{n_-}{H}\phantom{00.},\\
&\fl\Sigma_2\,=\,\frac{\sigma_{2}}{H}\,,\quad\quad\Theta=\frac{w}{\sqrt{6} H}\,,\quad\quad N_{\times}=\frac{n_\times}{H}\phantom{00.},\nonumber\\
&\fl\Sigma_3\,=\,\frac{\sigma_{3}}{H}\,,\quad\quad\Xi_i=\frac{q_i}{3H^2}\,\phantom{0},\quad\quad\Sigma^2=\frac{\sigma_{ab}\sigma^{ab}}{6 H^2}.\nonumber
\end{eqnarray}
where $H$ is the Hubble parameter.
In this way the equations of motion become an autonomous system of differential equations and all equilibrium points will represent self-similar cosmologies. The above definitions differ slightly from other authors (e.g. \cite{coley05, bok:EllisWainwright}), since we decompose such that
\label{App:Decomp}
\begin{equation}
x_{ab}=
\left(\begin{array}{ccc}
-2x_+      & \sqrt{3}x_2&\sqrt{3}x_3\\
\sqrt{3}x_2     & x_++\sqrt{3}x_-&\sqrt{3}x_\times\\
\sqrt{3}x_3     & \sqrt{3}x_\times&x_+-\sqrt{3}x_-\\
\end{array}\right),
\end{equation}
where $x_{ab}$ is one of the trace-less matrices $n_{ab}$ or $\sigma_{ab}$ (their normalized equivalents $N_{ab}$ and $\Sigma_{ab}$ have the same structure). Note that $n_{1b}=0$ (for all $b$) for the considered Bianchi type I-VII$_h$.

We align our frame such that the basis vectors $\mathbf{e}_A$ (where $A=\{2,3\}$) are aligned with the orbits of the $G_{2}$ subgroup permitted by the isometry group in the Solvable Bianchi types. This 1+1+2 split of space-time effectually fixes the two (expansion-normalized) rotations $R_2$ and $R_3$, but leaves a rotational gauge freedom $R_1$, as discussed in Section~\ref{Sec:GometryAndFrame}. More specifically, $R_1$ is the rotation of the frame around the $\mathbf{e}_1$-axis), which is orthogonal to the orbits of the $G_{2}$ subgroup. Taking the angle $\phi$ to be constant on the orbits of $G_{2}$ the (expansion-normalized) local angular velocity $R_{1}$ of a Fermi-propagated axis with respect to the triad $\bf{e}_{a}$ is given as
\begin{equation}
R_1\,=\phi'.
\end{equation}
Following~\cite{coley05}, we leave this gauge-freedom in the equations. Finally note the definitions
\begin{eqnarray}
\label{complexVar}
\eqalign{
\mathbf{N}_\Delta=N_-+iN_\times\,,\quad\quad &\mathbf{V}_c=V_2+iV_3\,,\\
\mathbf{\Sigma}_\Delta=\Sigma_-+i\Sigma_\times\,,\quad\quad &\mathbf{\Sigma}_1=\Sigma_2+i\Sigma_3\,.}
\end{eqnarray}
\newpage
\section{Dynamical system}
\label{App:dynsys}
The general system of equations before gauge choice is in \cite[Sec~4]{normann20a} derived from the general equations obtained in \cite[Sec~5]{normann18}. The resulting system of equations, which is given in the following, cover the Bianchi type A invariant sets $\mathcal{B}$(VI$_0$) and $\mathcal{B}$(VII$_0$) alongside the Bianchi type B invariant sets $\mathcal{B}$(IV), $\mathcal{B}$(VI$_h$) (where $h\,\neq\,-1\cup -1/9$) and $\mathcal{B}$(VII$_h$). These are the sets for which the curvature can be used to define the spatial frame unambiguously (once the gauge is chosen).
\begin{eqnarray}
&\fl\textit{j}\textrm{-form eq.s }\quad\quad\cases{\label{FluidEqsOpt1a}
V_1'=\left(q+2\Sigma_+\right)V_1,\\
\Theta'=(q-2)\Theta-2AV_1\,,}&\\
&\fl\textrm{Einst. eq.s }\quad\phantom{00}\cases{\label{EinstEqOpt1a}
\Sigma_-'=(q-2)\Sigma_-+2R_1\Sigma_\times+2 (A N_\times -N_- N_+)\\
\Sigma_\times'=(q-2)\Sigma_\times-2R_1\Sigma_--2 (A N_- +N_\times N_+)\\
\Sigma_+'=\left(q-2\right)\Sigma_+-2\left(N_-^2+N_\times^2+V_1^2\right)\,,}&\\
&\fl\textrm{En. cons. }\quad\quad\phantom{0}
\cases{\label{EnConsOpt1a}
\Omega_{\rm pf}'=2\left(q+1-\frac{3}{2}\gamma \right)\Omega_{\rm pf}\,,}&\\
&\fl\textrm{Jacobi id. }\quad\quad\phantom{.}
\cases{\label{JacId2Opt1a}
N_-'=\left(q+2\Sigma_+\right)N_-+2(R_1 N_\times+\Sigma_- N_+)\,,\\
N_\times'=\left(q+2\Sigma_+\right)N_\times-2(R_1 N_--\Sigma_\times N_+)\,,\\
N_+'=\left(q+2\Sigma_+\right)N_++6\left(\Sigma_-N_-+\Sigma_\times N_\times\right)\,,\\
A'=\left(q+2\Sigma_+\right)A.}&
\end{eqnarray}
Here $(^{'})$ represents derivative w.r.t. dynamical time $\tau$\setcounter{footnote}{0}\footnote{The dynamical time $\tau$ is defined such that ${\rm d}\tau/{\rm d}t=1/H$.}. The system of equations must obey the following constraints.
\begin{eqnarray}
&\fl\label{Constr1Option1a} C_1=1-\Omega_{\rm pf}-\Sigma_{+}^2-\Sigma_-^2-\Sigma_\times^2-\Theta^2-V_1^2-A^2-N_-^2-N_\times^2=0\,,\\
&\fl\label{Constr2Option1a} C_2=\Theta V_1-A\Sigma_+-N_\times\Sigma_-+N_-\Sigma_\times=0\,.
\end{eqnarray}
The invariant sets $\mathcal{B}$(VI$_h$) and $\mathcal{B}$(VII$_{h})$ has a group parameter $h$ defined through
\begin{equation}
\label{h}
A^2+h\left(3\abs{\mathbf{N}_\Delta}^2-N_+^2\right)=0. 
\end{equation}
\paragraph{Useful observation:} Also note that following directly from the system of equations above is the result that $V_1/A$ is a constant of motion;
\begin{equation}
\label{useful}
\left(\frac{V_1}{A}\right)'=0.
\end{equation}
Also note that $q$ may be expressed as
\vspace{10pt}
\begin{equation}
\label{q}\fl
q=2\Sigma^2+\frac 12({3}\gamma -2)\Omega_{\rm pf}+2\Theta^2-\Omega_\Lambda,\quad\textrm{where}\quad\Sigma^2\equiv\Sigma_{+}^2+\abs{\mathbf{\Sigma}_{\Delta}}^2+\abs{\mathbf{\Sigma}_1}^2.
\end{equation}

\subsection*{Choosing gauge}
\label{App:Gauge}
Since there is gauge freedom left in the equations, one may either choose to construct gauge independent variables, and cast the dynamical system in these variables instead. Alternatively, one may choose a particular gauge, and study the system in this gauge \cite{coley05}. Our analysis relies on the latter approach, and makes use of the following two gauges.
\begin{itemize}
\item Use the gauge freedom to diagonalize  $N_{ab}$. This means we let $N_+=\sqrt{3}\alpha\Re\{\mathbf{N}_\Delta\}$, by appropriately choosing $R_1$. We find 
\begin{equation*}
R_1=\sqrt{3}\alpha \Sigma_\times\phantom{00000}\textrm{and}\phantom{00000}N_+=\sqrt{3}\alpha N_-\quad\quad\quad\quad\textbf{($N_-$\,-\,gauge).}
\end{equation*}
for some function $\alpha$. If we use our remaining freedom (choosing $\phi_1(\tau=0)$) to say that $N_\times(\tau=0)=0$, then $N_\times$ will remain zero. Such a choice is possible, and we refer the reader to our previous works for more details.
\item A second choice that proves useful whenever $N_\times=N_+=0$ (e.g. Wonderland) is
\begin{equation*}
R_1=0.\quad\quad\quad\quad\quad\quad\quad\quad\quad\quad\quad\quad\quad\quad\quad\quad\quad\quad\quad\quad\textbf{($F$\,-\,gauge).}
\end{equation*}
In this case we should keep in mind that we have a constant gauge freedom left (namely $\phi_1(\tau=0)$).
\end{itemize} 

\newpage
\section*{References}
\bibliographystyle{JHEP}
\bibliography{BianchiReferences}

\end{document}